\documentclass[aps,prb,twocolumn,superscriptaddress,longbibliography]{revtex4-1}

\usepackage{amsmath}
\usepackage{amssymb}
\usepackage{natbib}
\usepackage{graphicx}
\usepackage{bm,hyperref}
\usepackage{xcolor}
\usepackage[normalem]{ulem}

\newcommand{\ool}[1]{\overline{\overline{#1}}}

\def\be{\begin{equation}}
\def\ee{{\bm e}}
\def\ba{\begin{eqnarray}}
\def\ea{\end{eqnarray}}
\def\rr{{\bm r}}

\def\w{\omega}
\def\W{\Omega}

\def\p{\partial}

\def\pp{{\bm p }}
\def\rr{{\bm r}}
\def\qq{\bm{q}}

\def\EE{\bm{\mathcal{E}}}
\def\BB{\bm{B}}

\def\d{\delta}

\def\bra{\langle}
\def\ket{\rangle}

\begin{document}
\title{Anomaly-induced sound absorption in Weyl semimetals}

\date{\today}
\author{Ohad Antebi}
\affiliation{Raymond and Beverly Sackler School of Physics and Astronomy,
Tel-Aviv University, Tel-Aviv 69978, Israel}
\author{D. A. Pesin}
\affiliation{Department of Physics, University of Virginia, Charlottesville, VA 22904, USA}
\author{A. V. Andreev}
\affiliation{Skolkovo Institute of  Science  and  Technology,  Moscow,  143026,  Russia}
\affiliation{Department of Physics, University of Washington, Seattle, Washington 98195, USA}
\affiliation{ L. D. Landau Institute for Theoretical Physics, Moscow, 119334, Russia}
\author{Roni Ilan}
\affiliation{Raymond and Beverly Sackler School of Physics and Astronomy,
Tel-Aviv University, Tel-Aviv 69978, Israel}
\begin{abstract}
We develop a semiclassical theory of sound absorption in Weyl semimetals in magnetic fields. We focus on the contribution to the absorption that stems from the existence of Berry monopoles in the band structure of such materials, or, equivalently, due to the chiral anomaly and chiral magnetic effect. Sound absorption is shown to come primarily from the motion of Weyl nodes in energy space, associated with the propagation of a sound wave. We argue that acoustic magneto-chiral dichroism, which occurs when absorption of sound is different for opposite propagation directions, and for opposite directions of the magnetic field, can be a definitive probe of band topology. The part of the monopole-related sound magnetoabsorption that is even in the magnetic field is negative in time-reversal Weyl materials. The difference in the sign of the effect as compared to the positive anomaly-related transport magnetoconductance stems from the existence of valley electrochemical imbalances without magnetic field in the sound propagation problem.  In centrosymmetric Weyl semimetals with few Weyl nodes, the magneto-absorption is negative at low frequencies, but can change sign with increasing frequency. 
\end{abstract}
\maketitle

\section{Introduction}
The classification of band-structures of crystalline materials according to their topological features relies  heavily on the concept of the Berry curvature\cite{berry1984phase,Xiao2010b}. The most  robust aspects of such a classification are related to the existence of topological invariants \cite{thouless1982quantized, avron1983homotopy, simon1983holonomy, kohmoto1985topological,VolovikBook}, which are defined as the flux of the Berry curvature through two-dimensional momentum space manifolds, and assume quantized values known as Chern numbers. For instance, the Chern number defined as the Berry curvature flux through the two-dimensional Brillouin zone defines the integer Hall conductivity of two-dimensional insulators\cite{bernevig2013topological}, and the Chern number that corresponds to the quantized Berry curvature flux through two-dimensional Fermi surfaces embedded in a three-dimensional Brillouin zone distinguishes topological (``Weyl'') and trivial metals \cite{Herring,murakami2007phase, wan2011topological, armitage2018weyl}. The physically observable manifestations of this quantized Berry flux in the three-dimensional topological metals that is the main focus of this work. In particular, we study how the existence of the Berry monopoles affects propagation of sound in Weyl metals.

The ongoing search for measurable quantities affected by topological invariants aims at linking the topological properties of a system to its physical responses. Over the past several decades, responses related to topology have been successfully identified both experimentally and theoretically in insulators. A non-exhaustive list of prominent examples of mostly electromagnetic nature includes the integer quantum Hall effect in two dimensional gases\cite{thouless1982quantized}, quantum anomalous Hall effect in topological insulators\cite{onoda2002topological, jungwirth2002ahe, Haldane2004,wang2007fermi, Nagaosa2010a, qi2006topological}, quantum spin Hall effect\cite{Kane2005b,bernevig2006quantum, qi2011topological, konig2007quantum, roth2009nonlocal, qi2006topological}, and quantized Kerr and Faraday effects on topological surfaces \cite{qi2008TFT, tse2010giant, tse2011magneto, jenkins2010kerr, aguilar2012terahertz, wu2016quantized}.

In the case of metals, the identification of low-frequency responses related to topology is complicated by their very metallicity, which usually means the part of the response related in some way to the topology has to be discerned from some mundane aspects of the physics of metals.

This implies that the search for effects related to topology should concentrate on either the magnitude or sign of an effect, setting it apart from the same phenomenon in conventional metals, or its dependence on parameters, like temperature, or frequency. The majority of proposals devoted to the identification of Berry monopoles in Weyl metals have followed this route: the existence of monopoles has been inferred from the temperature and frequency dependence of conductivity\cite{Hosur2012}, via the magnitude and sign of the longitudinal magnetoresistance\cite{Nielsen1983,SonSpivak2013}, in optical activity\cite{MaPesin2015,Zhong2015}, nonlocal transport\cite{ParameswaranPesin2014}, and in non-linear optical\cite{Juan2017cpge,Konig2019ghe} and transport properties\cite{NandyPesin2020}. Of these examples, only the circular photogalvanic effect \cite{Juan2017cpge} has been associated with some notion of quantization due to the existence of monopoles, which still can be masked by bandstructure and disorder scattering effects. 

In this work, we address the magnetic field dependence of acoustic attenuation in multivalley metallic systems as a diagnostic tool for the topology of their band structure. Specifically, we focus on the possibility to infer the existence of Berry monopoles in the band structure, which appear due to the presence of linear non-degenerate band crossings, from sound attenuation measurements. We show that the chiral anomaly\cite{Nielsen1983} and the chiral magnetic effect\cite{vilenkin1980equilibrium, fukushima2008chiral}, which exist only in the presence of Berry monopoles, lead to the acoustic magneto-chiral dichroism. This phenomenon is defined as nonreciprocal absorption of sound waves, being different for opposite orientations of the magnetic field, or opposite propagation directions. Hence acoustic magneto-chiral dichroism, being a quantum-mechanical effect with no classical counterpart in Drude-type transport, can be a smoking gun signature of the Berry monopoles in the band structure. We also show that there can be a significant magnetoabsorption response, which is even with respect to the magnetic field or sound propagation direction inversion, related to the existence of Berry monopoles in Weyl semimetals. Because of the conduction electrons screening of (pseudo)electric fields arising due to an acoustic perturbation, the mechanisms of acoustic magnetoabsorption turn out to be essentially unrelated to that of the negative longitudinal magnetoresistance, and are discussed in detail below.

The importance of the chiral anomaly and chiral magnetic effect for sound absorption in Weyl systems has been previously emphasized in Refs.~\onlinecite{pikulinfranz2016,andreevspivak2016}. Here, we develop the theory of sound magneto-absorption in Weyl metals based on the semiclassical transport equations \cite{Sundaram1999}, taking into account charge neutrality. We also note that conceptually the present study differs from the previous ones related to the transport and electromagnetic responses in that sound propagation through a material couples to its electronic subsystem through the modification of the band structure, rather than exciting electrons in a fixed band structure. This forces one to reconsider the phenomena traditionally related to the Berry monopoles - the chiral anomaly and chiral magnetic effect - for monopoles that have dynamics in the momentum or energy spaces.

The rest of the paper is organized as follows. Section~\ref{sec:qualitative} contains qualitative considerations that outline the physics of sound absorption in Weyl materials. Section~\ref{sec:calculations} describes the solution of the transport equations in Weyl semimetals in the presence of an acoustic perturbation and a magnetic field, and includes a discussion of restrictions imposed on the obtained results by crystalline symmetry. Section~\ref{sec:discussion} is devoted to a discussion of the obtained results, as well as contracting our results with those existing in the literature.  Section~\ref{sec:conclusions} contains concluding remarks.

\section{Qualitative considerations}\label{sec:qualitative}

\subsection{Parameter regimes}\label{sec:regimes}
\begin{figure}
  \centering
  \includegraphics[width=3.5in]{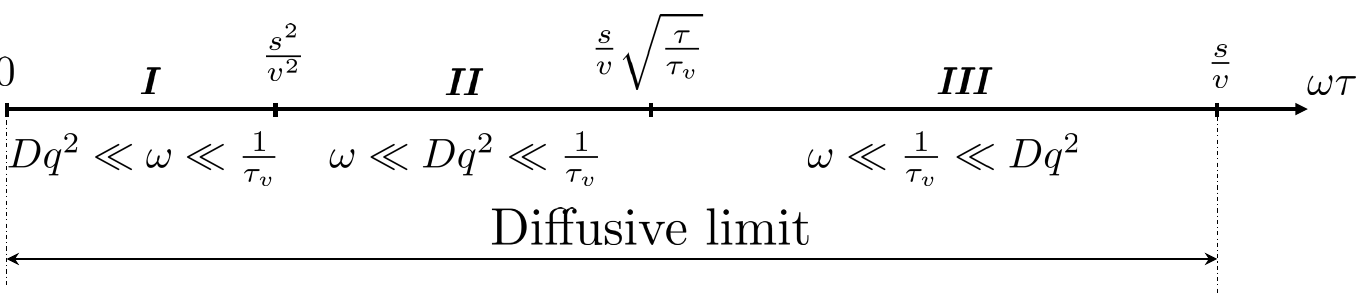}
  \caption{Frequency regimes (\textbf{\textit{I-III}}) with respect to relative magnitudes of $\omega$, $Dq^2$, and $1/\tau_v$. For definiteness, we assume the typical values of $\tau/\tau_v\sim 10^{-2}$(Ref.~\onlinecite{xiong2015evidence}),$\, s/v\sim 10^{-3}$, which implies $\sqrt{\tau/\tau_v}\gg s/v$. The widths of the parameter windows are not to scale. }\label{fig:frequencies}
\end{figure}

We start with a brief discussion of basic parameter regimes one can encounter in the problem of sound absorption in semimetals. The electronic response to a sound wave depends on the relation between the frequency of the wave, $\w$, its wave vector, $q=\omega/s$, where $s$ is the speed of sound, and the electronic transport mean free time, $\tau$, or mean free path, $\ell=v\tau$, where $v$ is the Fermi velocity. In multivalley semiconductors, another time scale of importance is the typical intervalley relaxation time, $\tau_v$. For typical values of parameters, inequalities $\omega\tau\ll1$ and $\tau/\tau_v\ll 1$ hold. However, due to the low speed of sound, the parameter $q\ell=\omega \tau v/s$ can be both small and large.

In this paper we confine ourselves to the practically important diffusive regime, $\omega\tau\ll1,\,q\ell\ll1$, or $\omega\tau\ll s/v$, in which electrons diffuse on the scale of the sound wavelength, which makes the response of electrons local in space. Electron diffusion introduces another time scale -- the time to diffuse across the wavelength of the sound wave -- given by $1/Dq^2$, $D$ being the electron diffusion coefficient. Within the diffusive regime, one can identify three distinct frequency intervals by comparing the relative magnitudes of $\omega$, $Dq^2\sim \omega^2\tau v^2/s^2$, and $1/\tau_v$. These intervals are labeled as $\textbf{\textit{I-III}}$ in Fig.~\ref{fig:frequencies}. Within intervals $\textbf{\textit{I,II}}$, the sound absorption is dominated by intervalley relaxation processes, while in interval $\textbf{\textit{III}}$ it is dominated by electron diffusion. 
It is worth keeping this information in mind while going through the rest of the paper.
We defer the detailed discussion of sound absorption in each frequency interval till Section~\ref{sec:mechanisms}.

An important characteristic of typical Weyl materials, relevant for sound absorption, is their considerable doping level, seen in most experimentally available samples. Therefore, charge density perturbations are screened: a sample with $N_v$ Weyl points with a typical Fermi energy $E_F$ would have an inverse screening length, $\kappa$, of order
\begin{align}
\kappa^2\sim\frac{N_v e^2}{\varepsilon\hbar v_F}\frac{E_F^2}{\hbar^2 v^2},
\end{align}
where $\varepsilon$ is the background dielectric permittivity of the material. Even at the edge of the diffusive regime, $q\ell\sim 1$, one has $\kappa/q\sim E_F\tau/\hbar\gg1$. Given the large number of Weyl points in typical materials\cite{weng2015weyl, xu2015discovery, yang2015weyl, lv2015experimental}, as well as possible trivial partially filled bands, it is clear that Weyl semimetals should be treated as good metals with regard to  absorption in the diffusive regime: electroneutrality must be maintained in the presence of a sound wave.

However, the Weyl semimetal is by definition a multivalley system. Unless all valleys are related by symmetry, and are identical with respect to their charge response, only the total local charge density is pinned by electroneutrality, while individual valley densities may develop nonequilibrium components that spread diffusively. This distinguishes the case of a Weyl semimetal, or any other multivalley doped semimetal with non-equivalent valleys, from the single-valley metal one, in which electroneutrality precludes any density diffusion.\cite{Akhiezer1957,AbrikosovBook}

\subsection{Mechanisms of sound (magneto-)absorption in Weyl semimetals}\label{sec:mechanisms}

The electronic contribution to the sound attenuation in any semimetal may be expressed in terms of the entropy production generated by a sound wave interacting with the electronic subsystem. What distinguishes Weyl semimetals from the textbook case of sound absorption in metals~\cite{AbrikosovBook} is their multivalley character, and the existence of the chiral anomaly. As discussed in Section~\ref{sec:regimes}, the presence of multiple valleys enables a sound wave to produce density imbalances between valleys even if the total density is fixed by strong screening (electroneutrality). Such inter-valley density imbalances are relaxed by relatively slow intervalley scattering, and can make a substantial contribution to the entropy production, and hence the sound absorption. This mechanism of sound absorption is well known in multivalley semiconductors\cite{GantmakherBook}. The truly unique aspect of Weyl semimetals is the existence of the chiral anomaly and the chiral magnetic effect in the presence of a magnetic field, which enables additional carrier redistribution among the nodes, and thus contributes to sound magneto-absorption. The latter is the main focus of the present work. Below we discuss main qualitative features of generation and relaxation of electronic disequilibrium Weyl semimetals, starting with a discussion of relaxation processes that are responsible for the electronic part of sound absorption.

The entropy production in the presence of a sound wave is related to dissipative processes: intra- and intervalley scattering. In general, the dissipation is dominated by the fastest relaxation process that is able to relax electron density in a given valley. Therefore, it is clear from Fig.~\ref{fig:frequencies} that intervalley scattering dominates the entropy production in regimes $\textbf{\textit{I}}$ and $\textbf{\textit{II}}$, while it is not operational in regime $\textbf{\textit{III}}$. In other words, the intervalley relaxation dominates sound absorption for
\begin{align}\label{eq:lowomega}
  \w\tau\lesssim\sqrt{\frac{\tau}{\tau_v}}\frac{s}{v_F},
\end{align}
while diffusion is the leading dissipation mechanism in the opposite limit.

We now turn our attention to the question of how a sound wave generates deviations from equilibrium in the electronic system. A weak strain associated with a sound wave does not lead to the disappearance of Weyl nodes, but rather displaces the nodes in energy and momentum spaces\cite{PhysRevB.94.241405}, as well as deforms them. The nonequilibrium parts of the electronic distribution function associated with the deformation of the Weyl cone, \textit{e.g.} deformation of iso-energetic surfaces, or change in the density of states, are relaxed on intravalley scattering time scales, and are important for single-valley semimetals\cite{Akhiezer1957}. In multivalley semimetals, such processes can be neglected in the diffusive regime due to their high relaxation rate. Therefore, below we focus on the first group of perturbations, those that correspond to displacement of the nodes in energy and momentum spaces. 

\begin{figure}
  \centering
  \includegraphics[width=3.5in]{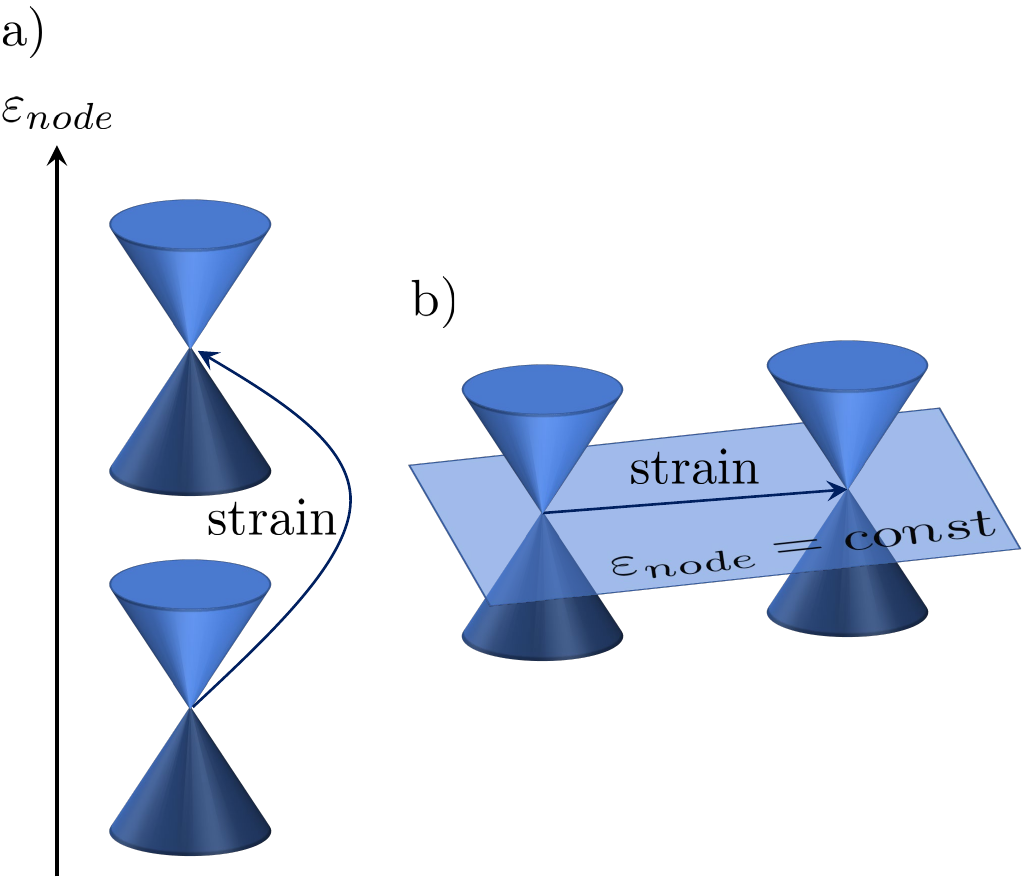}
  \caption{(Color online)  The two types of a Weyl node motion in a Weyl semimetal subject to the strain of an acoustic perturbation considered in this paper: a) ``scalar" perturbation: motion of a Weyl node in energy space without change of the nodal point location in the momentum space; b) ``vector" perturbation: motion of a Weyl node in momentum space along an iso-energetic surface. 
  }\label{fig:conemotion}
\end{figure}

These nodal displacements are illustrated in Fig.~\ref{fig:conemotion}. In particular, it is clear that nodal motions in energy and momentum space act like effective scalar and vector potentials of electromagnetism, respectively, from the point of view of a  low-energy theory of a single node (see Eq.\eqref{eq:acoustic_perturbation} below, and also  Ref.~\onlinecite{ilan2019pseudo} for a review). In what follows, we refer to the associated perturbations as ``scalar" and ``vector" ones, respectively. Both types of perturbations lead to the appearance of effective electric fields acting on Weyl fermions, which stem from spatial gradients of the nodal displacement in the energy space (scalar mechanism), or from the time dependence of the nodal displacement in the momentum space (vector mechanism). These perturbations are treated in detail in the next Section.

\section{Sound absorption in Weyl semimetals}\label{sec:calculations}

In this Section we present the technical part of this work: we describe the interaction of a Weyl semimetal with an acoustic wave, and calculate the corresponding rate of sound absorption. We also discuss the restrictions that crystalline symmetry imposes on the obtained results. The main results of this section are further discussed, and physically interpreted in Section~\ref{sec:discussion}. To avoid cluttering up the notation, we assume that the spectrum near each valley is isotropic. The generalization to the anisotropic case is trivial, but leads to cumbersome algebra. It does not affect the frequency or magnetic field dependence of the leading dissipation mechanism, described below. In accordance with this, the Hamiltonian of a single Weyl node is given by
\begin{align}\label{eq:Weyl_hamiltonian}
H_w&=E_w+v_w\bm \sigma\cdot\bm p.
\end{align}
Here $\pp$ is the quasimomentum counted from the position of the node in momentum space, and $v_w$ is the Fermi velocity, and $E_w$ is the position of the node in energy space.

Due to the pseudoscalar nature of $\bm\sigma\cdot\bm p$ product, Hamiltonian~\eqref{eq:Weyl_hamiltonian} possess chirality, which is given (for this simple isotropic case) by $\eta_w=\textrm{sign} (v_w)$. The chirality $\eta_w$ of a Weyl point is proportional to the Berry curvature flux through an isoenergetic surface of either of the two bands touching at the Weyl point\cite{Haldane2004}. Denoting $d\bm S$ to be the surface element on an isoenergetic surface in either the conduction or valence band near a particular Weyl point, with the convention that $d\bm S$ is directed along the outward and inward normals for the conduction and valence bands, respectively, the chirality of a Weyl point is given by
\begin{align}\label{eq:chirality}
  \eta_w=-\frac{1}{2\pi}\oint d\bm S\cdot\bm {\mathcal F},
\end{align}
 where $\bm {\mathcal F}$ is the Berry curvature of the band. Defined this way, the chirality is a property of  a Weyl point, and does not depend on which band - conduction or valence - is used to calculate the integral in Eq.~\eqref{eq:chirality}. As is clear from Eq.~\eqref{eq:chirality}, the chirality of a Weyl point is equivalent to the Weyl point's Berry monopole charge. 

We describe the interaction of a sound wave with the electronic subsystem through the deformation potential generated by its strain, $u_{ij}=(\p_i u_j+\p_j u_i)/2$, where $i$ is the Cartesian index, and $u_i$ is the displacement field associated with the wave. We disregard possible piezoelectric coupling in noncentrosymmetric crystals.  The main perturbation to the electronic Hamiltonian associated with the wave has the following general form:
\begin{align}\label{eq:acoustic_perturbation}
H_{\rm aw}&=(\lambda_{w,ij}+\bm\xi_{w,ij}\cdot\bm \sigma)u_{ij},
\end{align}
where $\lambda_{w, ij}$ and $\bm\xi_{w,ij}$ are the scalar and vector parts of the deformation potential in valley $w$: $\lambda_{w, ij}$ as well as the three components of vector $\bm\xi_{w,ij}$ are $3\times3$ matrices with respect to the Cartesian indices $i,j$. Summation over repeated indices is implied, and in what follows we will suppress explicit Cartesian indices, \textit{e.g.} $\lambda_{w, ij}u_{ij}\to \lambda_{w}u$. We note that the right hand side of Eq.~\eqref{eq:acoustic_perturbation} represents the matrix elements of the usual deformation potential Hamiltonian\cite{BirPikus1974} within the subspace spanned by the Weyl bands near a particular Weyl point. 
Furthermore, in Hamiltonian~\eqref{eq:acoustic_perturbation}, we neglected the momentum dependence of the deformation potentials. While it 
definitely exists in reality, only in single-valley metals it yields the leading term in the absorption\cite{Akhiezer1957,AbrikosovBook}. In the present multivalley case, momentum dependence of the deformation potential is usually of little importance.

The two perturbations described by the $\lambda$ and $\bm\xi$ terms in Eq.~\eqref{eq:acoustic_perturbation} describe motion of Weyl nodes in energy and momentum spaces, respectively, see Fig.~\ref{fig:conemotion}. Omitted $\bm p$-dependent terms correspond to momentum-space deformations at a single node.  In what follows, we will refer to the $\lambda$-term as the scalar perturbation, while to $\xi$ as the vector perturbation.  These give rise to effective (pseudo-)electric fields acting on carriers belonging to a particular node, much like the usual scalar and vector potentials in conventional electrodynamics. The magnitudes of the pseudoelectric fields stemming from the scalar, $\mathcal{E}^s$, and vector, $\mathcal{E}^v$ perturbations are
\begin{align}\label{eq:scalarvsvector}
e\mathcal{E}^{\rm s}_w&=-\bm\nabla\lambda_w u(\rr)\sim q \lambda_wu,\nonumber\\
e\mathcal{E}^{\rm v}_w&=\frac{1}{v_w}\partial_t \bm \xi_w u\sim \eta_w \frac{s}{\left|v_w\right|} q \xi_w u.
\end{align}
In these expressions, $e$ is the electron charge, $q$ is the wave vector of the acoustic wave, and we used the fact that for a harmonic perturbation $\p_t\sim \omega$. Note that the chirality of the node enters the field produced by the acoustic perturbation via the sign of the Fermi velocity.\cite{cortijo2015elastic} In typical crystals, $s/v_w\sim 10^{-3}$, and $\lambda_w\sim\xi_w$\cite{magnitudes_of_deformpotentials}, hence the vector perturbation produces much smaller pseudo-electric fields.  

\subsection{Absorption of sound due to scalar perturbations}\label{sec:scalarcase}

In this Section we consider absorption of sound in time-reversal invariant Weyl semimetals due to the scalar part of the deformation potential in the presence of a magnetic field. We focus on the magnetic field effects that arise due to the chiral anomaly, which is the key property associated with Weyl materials. We do not consider conventional mechanisms related to the intravalley cyclotron motion\cite{SteinbergSingleMagnetoattenuation}, which are identical to those of common multivalley semiconductors. 

In the diffusive regime, $q\ell\ll1$, the response of the electronic subsystem to an acoustic perturbation can be found from a macroscopic transport equation. In the case of a Weyl system, such an equation is represented by the diffusion equation augmented with additional terms related to the chiral anomaly and the chiral magnetic effect.

The transport equation for the change in the charge density, $n_w$, in valley $w$ can be most economically written in terms of the nonequilibrium part of the electrochemical potential for that valley, $\mu_w$:
\begin{align}\label{eq:elchempot}
    \mu_w=\lambda_wu+e\phi+n_w/\nu_w.
\end{align}
In this expression, $\lambda_wu$ plays the role of the valley-dependent electric potential, $e\phi$ is the usual electric potential due to the screening charges, and $n_w/\nu_w$ is the change in the chemical potential near a valley due to the acoustic wave.

For future convenience, we introduce the following notation for the averages over Weyl nodes weighted with the density of states:
\begin{align}\label{eq:weightedaverage}
    \ool O\equiv\frac{\sum_w\nu_w O_w}{\sum_w\nu_w}
\end{align}
Using these notations, the transport equation for $n_w$ can be written as follows:
\begin{widetext}
\begin{align}\label{eq:transport}
    \p_t n_w-\nu_w D_w\nabla^2\mu_w+\frac{e}{4\pi^2}\eta_w\bm B\cdot\bm\nabla\mu_w=
    -\frac{\nu_w}{\tau_v}\left(\mu_w-\ool \mu\right).
\end{align}
In the absence of a deformation potential, this transport equation was obtained in Ref.~\onlinecite{ParameswaranPesin2014}. For completeness, we derive it in Appendix~\ref{sec:entropy_appendix} taking account of the deformation potential. 
\end{widetext}

In Eq.~\eqref{eq:transport} the presence of the Berry monopoles manifests itself in the third (linear in the  $\bm B$-field) term on its left hand side. This term is a total spatial derivative, and combines both the divergence of the chiral magnetic effect current, and the expression for the chiral anomaly in the presence of a potential electric field.\cite{ParameswaranPesin2014} This term can be thought of as a divergence of the generalized valley-specific CME current,
\begin{align}\label{eq:CMEcurrent}
  \bm j^{\rm{cme}}_w\equiv \frac{e^2}{4\pi^2}\eta_w \mu_w \BB.
\end{align}
We reiterate that $\mu_w$ in Eq.~\eqref{eq:CMEcurrent} represents the total electrochemical potential in valley $w$, given by Eq.~\eqref{eq:elchempot}, and the generalized CME current embodies both the usual CME current driven by the chemical potential of a valley, as well as the effect of the chiral anomaly driven by a potential electric field. 

The intervalley collision integral on the right hand side of Eq.~\eqref{eq:transport} is written in the relaxation-time approximation. It is easy to show that this collision integral satisfies two basic physical requirements: it conserves the total particle density in all valleys, and it vanishes when all valleys have coincident electrochemical potentials.

In order to determine the potential of the screening electric field, $\phi$, Eq.~\eqref{eq:transport} must be supplemented with the Poisson equation. However, as explained in Section~\ref{sec:regimes}, electroneutrality is maintained during propagation of a sound wave. Therefore, we replace the Poisson equation with the electroneutrality condition\cite{Akhiezer1957}:
\begin{align}\label{eq:neutrality}
    \sum_w n_w(\rr,t)=0.
\end{align}
This substantially simplifies the solution.

Entropy production in the electronic subsystem determines the energy loss of the acoustic wave. In the present case, the entropy production comes from intravalley Joule heat as well as from intervalley scattering. Taking the standard route (see Appendix \ref{sec:entropy_appendix}), we obtain the following expression for the entropy, $S$, production:
\begin{align}\label{eq:entropy_production}
    T\dot S=\sum_w\int_\rr \nu_w D_w(\bm\nabla \mu_w)^2+\sum_w\int_\rr\nu_w\frac{1}{\tau_v}\left(\mu_w-\ool\mu\right)^2,
\end{align}
where $T$ is the sample temperature. The first term in the entropy production equations corresponds to intravalley diffusion, and the second one comes from intervalley scattering. Note that the electric potential $e\phi$, which is independent of the valley index, drops out from the entropy production term associated with the intervalley scattering.

Equations~\eqref{eq:transport},~\eqref{eq:neutrality}, and~\eqref{eq:entropy_production} constitute a full set of equations required to determine the electronic contribution to sound absorption in the diffusive regime.

The system of algebraic Eqs.~\eqref{eq:transport},~\eqref{eq:neutrality}, and~\eqref{eq:entropy_production} admits a straightforward solution procedure. First, one determines the electrochemical potentials in all valleys, $\mu_w$ from Eq.~\eqref{eq:transport}:
\begin{align}\label{eq:chempot}
    \mu_w=\frac{-i\w(\lambda_w u+e\phi)+\frac{1}{\tau_v}\ool\mu}{-i\w+D_wq^2+i\W_w\eta_w+\frac{1}{\tau_v}}.
\end{align}
where
\begin{align}\label{eq:Omega}
    \W_w=\frac{e}{4\pi^2\nu_w}\bm q\bm B.
\end{align}
Using this equation, $\ool\mu$ can be calculated self-consistently, such that $e\phi$ is the only unknown in Eq.~\eqref{eq:chempot}. The screening electric potential is determined by imposing the electroneutrality condition~\eqref{eq:neutrality} using
\begin{align}
    n_w=\nu_w\frac{-(D_wq^2+i\W_w\eta_w+\frac{1}{\tau_v})(\lambda_w u+e\phi)+\frac{1}{\tau_v}\ool\mu}{-i\w+D_wq^2+i\W_w\eta_w+\frac{1}{\tau_v}},
\end{align}
which follows from Eqs.~\eqref{eq:transport} and \eqref{eq:chempot}. Once $\ool\mu$ and $e\phi$ have been determined, one can calculate the entropy production (and the absorption coefficient) from Eq.~\eqref{eq:entropy_production}.

In practice, the procedure outlined above is quite cumbersome for arbitrary frequencies, magnetic fields, and valley characteristics ($\nu_w,D_w$). Therefore, in what follows we assume that the variation of the deformation potential among Weyl nodes is the leading cause of sound absorption, the differences in the diffusion constants and densities of states being relatively small. This is often the situation encountered in practice~\cite{GantmakherBook}. Hence, we set $D_w\to D,\nu_w\to \nu$, thus $\Omega_w\to\Omega$, for all valleys, but keep valley-dependent deformation potentials, $\lambda_w$, in Eqs.~\eqref{eq:transport},~\eqref{eq:neutrality}, and~\eqref{eq:entropy_production}.

As a result, we obtain the following expression for the value of the screening potential:
\begin{align}\label{eq:elpot}
  e\phi&=-\ool\lambda u+e\Phi,
\end{align}
where
\begin{align}\label{eq:bigphi}
  e\Phi&=-\frac{\w\Omega}{Dq^2(-i\w+Dq^2+\frac{1}{\tau_v}+\frac{\Omega^2}{Dq^2})}\ool{\eta\lambda}u.
\end{align}
Physically, the screening potential is set to nullify the total longitudinal electric current caused by effective electromagnetic fields accompanying the sound wave propagation. This prevents the total current from having a spatial divergence, which would otherwise generate a net local charge accumulation. The first term on the right hand side of Eq.~\eqref{eq:elpot} nullifies the usual diffusive current\cite{Akhiezer1957,andreevspivak2016}. The second term - $e\Phi$ from Eq.~\eqref{eq:bigphi} - stems from the existence of the CME current. This contribution to the screening potential is unique to Weyl semimetals, and is one of the results of the present work. It will be shown below that $e\Phi$ is the source of the leading anomaly-induced magnetic field dependence of sound absorption. 

Assuming the amplitude of the sound wave to be constant throughout the crystal, the total entropy production rate averaged over an oscillation period can be written as 
\begin{widetext}
\begin{align}\label{eq:scalartotal}
 \langle T\dot{S}^{\rm s}(\bm B)\rangle_{\rm osc}&=\frac{\nu V}{2}\sum_w\left[\frac1{\tau_v}\frac{|i\w\left(\lambda_w-\ool{\lambda }\right)u+(Dq^2+i\Omega\eta_w)e\Phi|^2}
 {(\omega-\Omega\eta_w)^2+\frac{1}{\tau^2_q}}
 +Dq^2 \frac{|i\w\left(\lambda_w-\ool{\lambda }\right)u+(i\w-1/\tau_v)e\Phi|^2}
 {(\omega-\Omega\eta_w)^2+\frac{1}{\tau^2_q}}\right],
\end{align}
where $\langle\ldots\rangle_{osc}$ stands for  averaging over an oscillation period, $V$ is the volume of the system, and 
\begin{align}\label{eq:tauq}
  \frac{1}{\tau_q}=\frac1{\tau_v}+Dq^2
\end{align}
is the total relaxation rate for individual valley densities, which includes contributions from both intervalley scattering and diffusion.
\end{widetext}

The entropy production rate of Eq.~\eqref{eq:scalartotal} is one of the main results of this paper. In noncentrosymmetric crystals, it contains both even and odd in $B$-field parts. The even part describes the usual magnetoabsorption, and the odd one describes the magneto-chiral dichroism of acoustic waves: the absorption rate is different for opposite directions of propagation. These parts are discussed below. 

We emphasize that within the present treatment the magnetic field dependence of the entropy production in Eq.~\eqref{eq:scalartotal} stems only from the existence of the Berry curvature monopoles, which manifest themselves through the chiral anomaly, and the chiral magnetic effect. This is not the entire magnetic field dependence: there is a variety of contributions to the magneto-absorption, for instance, stemming from the magnetic field dependence of the quantities entering in Eq.~\eqref{eq:scalartotal} (\textit{e.g.} the density of states\cite{berry_dos_correction}, the diffusion coefficient, the intervalley relaxation time\cite{GantmakherBook}, etc.). All such corrections are governed by the usual $\omega_c\tau$ parameter, in which $\omega_c$ is the cyclotron frequency at the Fermi level, and $\tau$ is the intravalley scattering time.

\begin{figure}
  \centering
  \includegraphics[width=3.0in]{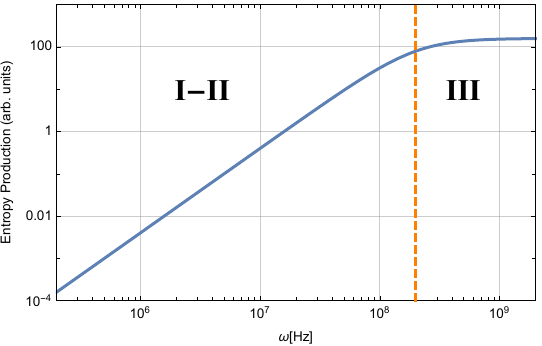}
  \caption{
  Log-log plot of the entropy production as a function of frequency for the scalar mechanism without magnetic field. We use velocities and timescales that reflect typical values: $s = 10^{3}\frac{m}{s},v=10^6\frac{m}{s},\tau = 0.5\times10^{-12}s,\tau_{v}=0.5\times10^{-10}s$.
  The maximal frequency corresponds to $\omega\tau=s/v$ which is the highest frequency within the diffusive regime. The crossover between intervalley dominated ($\textbf{\textit{I}}-\textbf{\textit{II}}$) and diffusion dominated ($\textbf{\textit{III}}$) regimes of Fig.~\ref{fig:frequencies} is shown by a dashed line.
  }\label{fig:scalar_zero_field}
\end{figure}

Eq.~\eqref{eq:scalartotal} allows one to study various regimes of sound absorption. In the absence of the magnetic field, a  Weyl metal is completely analogous to a multivalley doped semiconductor as far as the sound absorption goes, and we obtain the standard expression for the dissipation rate~\cite{GantmakherBook}:
\begin{align}\label{eq:scalarnoB}
  \langle T\dot{S}^{\rm s}(0)\rangle_{\rm osc}&=\frac{\nu N_v V}{2}\frac{\w^2\tau_q}
 {\omega^2\tau_q^2+1}\ool{|(\lambda-\ool{\lambda })u|^2}.
\end{align}
The dissipation rate scales quadratically with frequency in  intervals $\textbf{\textit{I-II}}$, in which it is dominated by intervalley relaxation processes, and reaches a plateau in interval \textbf{\textit{III}}, where it is determined by intravalley diffusion. It is plotted in Fig.~\ref{fig:scalar_zero_field}. We note that the factor $\omega\tau_q$ satisfies  $\omega\tau_q\ll1$ in the entire range of validity of the present theory (see Fig.~\ref{fig:frequencies}), hence we will neglect it as compared to unity in what follows. 

A compact expression for the magnetoabsorption at small $B$-fields for small frequencies can also be obtained from Eq.~\eqref{eq:scalartotal}. For later convenience, it is useful to express it in terms of the cyclotron frequency, $\w_c=e B v^2/\mu$, where $\mu$ is the doping of a Weyl point away from the energy of the Weyl node, which is assumed to be the same for all Weyl points for simplicity. Then $\Omega$ in Eq.~\eqref{eq:scalartotal} is given by $\W=\frac{v}{s}\frac{\w_c |\w|}{2\mu} \bm e_{\bm B}\cdot \bm e_{\bm q}$, where $\bm e_{\bm B},\,\bm e_{\bm q}$ are the unit vectors along the directions of $\bm B$ and $\bm q$, respectively, see Eq.~\eqref{eq:Omega}. 

We start with the odd in the $B$-field part of the entropy production, which is linear in $B$ at small magnetic fields. We restrict ourselves to the experimentally relevant $\omega\tau_q\ll1$ regime. Expanding Eq.~\eqref{eq:scalartotal} in $\Omega\propto B$, and keeping the leading term, we obtain the following expression for the odd part of entropy production rate: 
\begin{align}\label{eq:entropyproduction_scalar_odd}
    \langle T\d\dot{S}^{\rm {s,odd}}(\bm B)\rangle_{\rm osc}=\frac{\nu N_v V}{2} (\omega\tau_q)^3\frac{v}{s}\frac{\w_c |\w|}{\mu} \bm e_{\bm B}\cdot \bm e_{\bm q}\ool{\eta\left|(\lambda-\ool{\lambda })u\right|^2}.
\end{align}
Note that expression~\eqref{eq:entropyproduction_scalar_odd} is odd in the wave vector of the sound wave, as dictated by the Onsager symmetry~\cite{Melrose}: $T\dot{S}(\omega,\qq,\BB)=T\dot{S}(-\omega,\qq,-\BB)=T\dot{S}(\omega,-\qq,-\BB)$. 

Turning to the even in the $B$-field part of the entropy production in Eq.~\eqref{eq:scalartotal}, we note that at small magnetic fields it is quadratic in $B$. To the leading order, at small-frequencies and small fields it is given by
\begin{align}\label{eq:smallomegaScalarB}
   \langle T\d\dot{S}^{\rm {s,even}}(\bm B)\rangle_{\rm osc}
  =-\frac{\nu N_v V}{8}\omega^2\tau_v\frac{v^2\tau_v}{D} \frac{\w_c^2}{\mu^2}(\bm e_{\bm B}\cdot\bm e_{\bm q})^2|\ool{\eta\lambda}u|^2.
\end{align}

The leading contribution to the even in B-field part of the magnetoabsorption comes from the magnetic field dependence of the screening potential ($e\Phi$ term in Eq.~\eqref{eq:elpot}, see also Eq.~\eqref{eq:bigphi}), and appears at $O(\w^2)$ order. The origin of this negative magnetoabsorption is elaborated upon in Section~\ref{sec:previouswork}.

The existence of contributions to entropy production given by Eqs.~\eqref{eq:entropyproduction_scalar_odd} and~\eqref{eq:smallomegaScalarB} relies on the valley sums in their right hand sides being nonzero. This imposes symmetry constraints on crystals in which these contributions exist. These symmetry constraints are discussed in Section~\ref{sec:symmetry}. Here we only mention that in crystals with symmetry groups such that $e\Phi\propto \ool{\eta \lambda}u=0$, the leading anomaly-related contribution to the magnetoabsorption appears at $O(\omega^4)$ order, and reads 
\begin{align}\label{eq:subleading entropy production}
 \langle T\dot{S}^{\rm s}(\bm B)\rangle_{\rm osc}&=-\frac{\nu N_v V}{8}\omega^4\tau^3_v\frac{v^2}{s^2} \frac{\w_c^2}{\mu^2}(\bm e_{\bm B}\cdot\bm e_{\bm q})^2\ool{|(\lambda-\ool{\lambda })u|^2}.
\end{align}

If the contributions to magnetoabsorption of Eqs.~\eqref{eq:smallomegaScalarB} and Eq.~\eqref{eq:subleading entropy production} are both present in a crystal of given symmetry, the latter is obviously suppressed at low frequencies. However, it becomes comparable to the former for $\omega\tau\gtrsim\sqrt{\frac{\tau}{\tau_v}}\frac{s}{v}$, which is in the crossover region between frequency intervals $\textbf{\textit{II}}$ and $\textbf{\textit{III}}$ of Fig.~\ref{fig:frequencies}.

\subsection{Absorption of sound due to vector perturbations in TR-invariant Weyl semimetals}\label{sec:vectorTR}

\begin{figure}[t]
  \centering
  \includegraphics[width=3.3in]{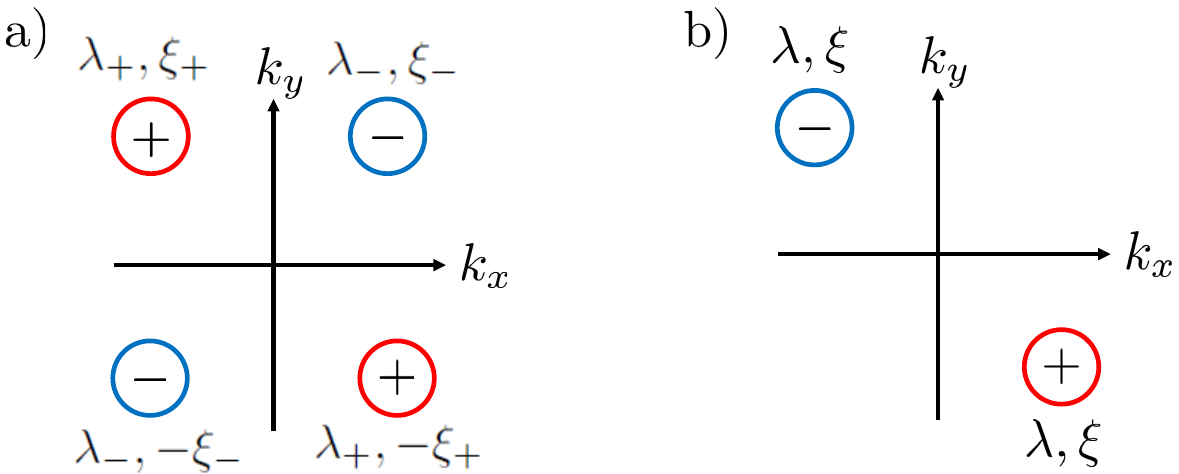}
  \caption{(Color online) Schematic Weyl node configuration of the simplest models of $(a)$ TR-invariant and $(b)$ centrosymmetric TR-breaking Weyl semimetals. The circles represent the Fermi surfaces of the Weyl nodes. The $\pm$ signs inside each circle denote the chirality of the node. The scalar ($\lambda$) and vector ($\xi$) deformation potentials of each node are shown. The nodal motion in momentum space induced by vector perturbations is governed by $\eta_w\xi_w$, and is opposite for nodes related by either TR or inversion symmetries in both models.
  }\label{fig:weylpoints}
\end{figure}

We have defined vector perturbations as those that displace the location of a Weyl node in momentum space without changing its energy, see Eq.~\eqref{eq:Weyl_hamiltonian}, and \eqref{eq:acoustic_perturbation}, as well as Fig.~\ref{fig:conemotion}. According to Eq.~\eqref{eq:scalarvsvector}, vector perturbations in general create much smaller effective electric fields as compared to the scalar case. However, as will become clear momentarily, vector perturbations do not disturb the total charge density in TR-invariant Weyl semimetals, hence are not susceptible to screening. Therefore, there might be a competition between vector and scalar perturations, which we explore in this Section.  We will see below that despite the lack of screening, vector perturbations yield a small contribution to overall attenuation rate. We keep the isotropic model adopted above, $\nu_w\to\nu, D_w\to D$, since it is definitely sufficient to estimate the relative importance of various absorption mechanisms.

First, we argue that local density disturbances produced by vector perturbations lead to very small contributions to entropy production, and can be neglected. To this end, we note that Weyl nodes related by the TR symmetry are constrained to stay at opposite quasimomenta. Their displacements (set by $\eta_w \bm \xi_w u$, as Eqs.~\eqref{eq:Weyl_hamiltonian} and \eqref{eq:acoustic_perturbation} show) are thus opposite, as illustrated in Fig.~\ref{fig:weylpoints}a. This implies that if nodes $w$ and $\overline w$ are related by the TR, then the effective electric fields in Eqs.\eqref{eq:scalarvsvector} are opposite, i.e. $e\mathcal{E}^{\rm v}_w=-e\mathcal{E}^{\rm v}_{\overline w}$. It then follows that these fields drive opposite Ohmic electric currents (with opposite divergences) in valleys $w$ and $\overline w$, which cannot produce any net local density perturbation. This statement would be invalid in the presence of tilt of electronic dispersion near a Weyl node. In that case a valley-specific conductivity tensor has a Hall component due to free carriers.~\cite{Steiner2017tilt} These valley Hall conductivities are of opposite sign for valleys related by the TR symmetry, and the corresponding Hall currents flow in the same direction, since the pseudoelectric fields are also opposite. However, it is straightforward to show that even a finite tilt leads to very weak effects, which do not change our conclusions~\cite{footnote_tilt}, so we assume that there is no tilt from here on. Further, pseudoelectric fields do not change the total charge density $n_w+n_{\overline w}$ in the two valleys in the presence of a magnetic field and the chiral anomaly. Indeed, since the chiralities of the Weyl nodes related by the TR are the same, while the corresponding pseudoelectric fields are opposite, the charge density near the Fermi level does not change because of the spectral flow between the valleys, because $\sum_w\eta_w \bm{\mathcal{E}}^{\mathrm{v}}_w\cdot \bm B=0$. Hence, we conclude that vector perturbations do not lead to the total electronic density changes in TR-invariant Weyl semimetals.

The fact that vector perturbations do not create density disturbances in TR-invariant Weyl semimetals simplifies the transport equation for this case. To write it, we need to discard the screening electric potential in Eq.~\eqref{eq:transport}, and use the expression for the effective electric field acting on an electron in valley $w$, given by (see Eqs.~\eqref{eq:Weyl_hamiltonian}, \eqref{eq:acoustic_perturbation}  and ~\eqref{eq:scalarvsvector})
\begin{align}\label{eq:vectorfield}
    e{\bm {\mathcal{E}}}_w^{\mathrm{v}}=\eta_w\frac{1}{v}\bm \xi_w \partial_t u.
\end{align}
The Dirac velocity $v$ is assumed to be the same in all valleys for simplicity. 

The transport equation for the density in valley $w$ then can be written as follows:
\begin{widetext}
\begin{align}\label{eq:transport_vector}
   \partial_t n_w+\nu D\bm \nabla\cdot\left(e\bm{\mathcal{E}}_w^{\mathrm{v}}-\frac{1}{\nu}\bm\nabla n_w\right)-\frac{e}{4\pi^2}\eta_w\bm B\cdot\left(e\bm{\mathcal{E}}_w^{\mathrm{v}}-\frac{1}{\nu}\bm\nabla n_w\right) =
   -\frac{1}{\tau_v}n_w.
\end{align}
\end{widetext}
As already mentioned, this equation is a direct analog of the transport equation for the scalar case, Eq.~\eqref{eq:transport}, in which $-\bm\nabla \mu_w$ is replaced according to $-\bm\nabla\mu_w\to e\bm{\mathcal{E}}_w^{\mathrm{v}}-\frac{1}{\nu}\bm\nabla n_w$ in the left hand side, while  $\mu_w\to\frac{1}{\nu}n_w$, $\ool\mu\to0$ in the collision integral on the right hand side. Then, the second term on the left hand side is the divergence of the usual transport current driven by the sum of a mechanical, $e\bm{\mathcal{E}}_w^{\mathrm{v}}$, and statistical, $-\frac{1}{\nu}\bm\nabla n_w$, forces. In turn, the third term on the left hand side of Eq.~\eqref{eq:transport_vector} consists of two parts:  the already familiar one with $\frac{1}{\nu}\BB \cdot \bm\nabla n_w$ describes the divergence of the CME current, and the term containing $\BB\cdot\EE_{w}^{\mathrm{v}}$ describes the spectral flow -- the chiral anomaly -- driven by the pseudoelectric fields due to the vector perturbation. Finally, since vector perturbations do not create any net local density perturbation (as explained above, we neglect possible effect of Weyl cone tilt), and do not cause motion of Weyl nodes in the energy space, the intervalley collision integral in the relaxation time approximation simply relaxes the nonequilibrium part of chemical potential near a given Weyl point. In our approximation of equal density of states, the nonequilibrium part of a valley chemical potential is equal to $n_w/\nu$, hence the form of the right hand side of Eq.\eqref{eq:transport_vector}.

The discussion of the preceding paragraph, combined with the understanding of Eq.~\eqref{eq:entropy_production} developed in Appendix~\ref{sec:entropy_appendix}, also makes it clear that the contribution of the vector perturbation to the entropy production rate (also denoted by superscript ``v'') is given by
\begin{align}\label{eq:entropy_production_vector_general}
    T\dot S^{\rm{v}}=\sum_w\int_\rr \nu D\left(e\bm{\mathcal{E}}_w^{\mathrm{v}}-\frac{1}{\nu}\bm \nabla n_w\right)^2+\sum_w\int_\rr\frac{n_w^2}{\nu\tau_v}.
\end{align}

In the isotropic model we consider, vector perturbations can be subdivided into two classes: longitudinal, $\bm{\mathcal{E}}_w\propto \bm q$, and transverse, $\bm{\mathcal{E}}_w\cdot\bm q=0$. In what follows we discuss the cases of transverse and longitudinal vector perturbations in TR-invariant Weyl semimetals. 

\subsubsection{Transverse vector perturbations}

We start with the dissipation due to the transverse vector perturbation in the $B=0$ case. In this case, the transverse nature of the electric field, ${\bm {\mathcal{E}}}_w\cdot \bm q=0$, ensures that it drives divergence-free electric currents, which do not perturb charge densities in all of the individual valleys. This implies that for $B=0$, the dissipation due to the transverse vector mechanism only comes from the standard Joule heating, given by the first term on the right hand side of Eq.~\eqref{eq:entropy_production_vector_general}. The entropy production averaged over an oscillation is given by (``v"-vector, ``t"-transverse)
\begin{align}\label{eq:vectortransversenoB}
    \langle T\dot S^{\rm{v},\rm{t}}(0)\rangle=\frac{\nu V}2  \sum_w  \frac{D \omega^2}{v^2} (\bm \xi_wu)^2.
\end{align}

To write this expression we set $n_w=0$ in Eq.~\eqref{eq:entropy_production_vector_general}, and used Eq.~\eqref{eq:vectorfield}.

Turning to the chiral-anomaly-induced $B$-field dependence of sound absorption due to the vector transverse mechanism, we note that for our isotropic model with valley-independent DoS, the corresponding contribution to the entropy production is similar to the usual positive magnetoconductance~\cite{Nielsen1983,SonSpivak2013}. Transverse vector perturbations produce density, and hence chemical potential, imbalances between valleys in the presence of a magnetic field. As is evident from transport equation~\eqref{eq:transport_vector}, the opposite effective electric fields in TR-related valleys with the same chirality produce spectral flow between such valleys, quite analogous to the usual case of the spectral flow between valleys with the opposite chirality in the presence of an external electric field and a transport electric field. This spectral flow changes individual valley densities, which leads to dissipation due to both Joule heat produced by currents driven by density gradients, as well as due to interavalley scattering:
\begin{align}\label{eq:vtB}
\bra T\d\dot S^{\rm{v,t}}(\bm B) \ket_{\rm{osc}}=\frac{\nu N_v V}{8}\w^2\tau_q\frac{\w_c^2}{\mu^2}\ool{(\bm e_{\bm B}\cdot\bm \xi u)^2}.
\end{align}
The presence of $\tau_q$ of Eq.~\eqref{eq:tauq} in this expression signals that individual valley densities relax both by intervalley scattering and diffusion. We used the electric field from Eq.~\eqref{eq:vectorfield} to write the above expression. It also should be noted that Eq.~\eqref{eq:vtB} is valid to the lowest (quadratic) order in $B$.

\subsubsection{Longitudinal vector perturbations}
Longitudinal vector perturbations are defined by $\bm \xi_w=\bm e_{\qq}\xi^q_w$. In this case, the forces acting on the electrons due to the pseudoelectric fields, and statistical forces due to chemical potential gradients are parallel to each other. Just like in the case of a scalar perturbation, this leads to the existence of an odd-in-magnetic field part in the entropy production, in addition to the usual magnetoabsorption that is quadratic in the magnetic field at small fields. The dissipation rate is obtained from Eqs.~\eqref{eq:transport_vector} and \eqref{eq:entropy_production_vector_general}, by switching to the Fourier space. We omit the tedious algebraic manipulations, and provide the final results for the dissipation rate in the absence of a magnetic field: 
\begin{widetext}
\begin{align}\label{eq:vectorlongitudinalnoB}
    \langle T\dot S^{\rm{v},\rm{l}}(0)\rangle=\frac{\nu N_v V}2  \frac{D\omega^2}{v^2}\frac{\tau_q}{\tau_v} \ool{|\xi^q u|^2},
\end{align}
as well as the odd-in-magnetic-field part of the entropy production,
\begin{align}\label{eq:entropyproduction_vector_odd}
    \langle T\d\dot{S}^{\rm {v,l,odd}}(\bm B)\rangle_{\rm osc}=\frac{\nu N_v V}{2}\frac{D}{v^2\tau_v} (\omega\tau_q)^3\frac{v}{s}\frac{\w_c |\w|}{\mu} \bm e_{\bm B}\cdot \bm e_{\bm q}\ool{\eta\left|\xi^qu\right|^2},
\end{align}
and the odd-in-magnetic-field part of the entropy production:
\begin{align}\label{eq:smallomegavectorB}
   \langle T\d\dot{S}^{\rm {v,l,even}}(\bm B)\rangle_{\rm osc}
  =\frac{\nu N_v V}{8}\frac{\tau^2_q}{\tau^2_v}\omega^2\tau_q \frac{\w_c^2}{\mu^2}(\bm e_{\bm B}\cdot\bm e_{\bm q})^2\ool{| \xi^q u|^2}.
\end{align}
\end{widetext}

\subsection{Absorption of sound in magnetic centrosymmetric Weyl semimetals}

To consider sound absorption in centrosymmetric Weyl metals with broken time-reversal invariance, we consider the minimal model of such a material. Such a model contains two Weyl nodes of opposite chirality, which are located at the same energy.

The inversion symmetry of the crystal places obvious restrictions on the scalar and vector parts of the deformation potential, which are illustrated in Fig.~\ref{fig:weylpoints}b. The scalar deformation potentials must coincide in the two valleys. As is clear from the discussion of the scalar mechanism of sound absorption in the noncentrosymmetric case, the average deformation potential gets screened out completely, hence the scalar mechanism is simply non-operational in the two-valley case.\cite{footnote_akhiezer} Further, for the case of a vector perturbation, the inversion symmetry restricts the effective electric fields to be opposite in the two valleys,
\begin{align}\label{eq:E5inversion}
  \bm{\mathcal{E}}^{\rm{v}}_{+}=-\bm{\mathcal{E}}^{\rm{v}}_{-}\equiv \bm{\mathcal{E}}_5.
\end{align}

The transport equation for this case is quite analogous to Eq.~\eqref{eq:transport_vector}, but one has to make account for the possibility of a finite screening potential, $\phi$, as explained below. Therefore, it is $\mu_w=e\phi+\frac{1}{\nu}n_w$ that enters into the transport equation:
\begin{widetext}
\begin{align}\label{eq:transporttwovalleys}
    \p_t n_w+\nu D\bm\nabla\cdot(e \bm{\mathcal{E}}^{\rm v}_w-\bm\nabla\mu_w)-\frac{e}{4\pi^2}\eta_w\bm B\cdot(e\bm{\mathcal{E}}^{\rm v}_w-\bm\nabla\mu_w)=
    -\frac{\nu}{\tau_v}\left(\mu_w-\ool \mu\right).
\end{align}
\end{widetext}
Unlike the case of TR-invariant crystals, in the present case the chiral anomaly driven by $\EE_{w}^{\rm v}$, which is described by the term containing $\BB\cdot\EE_{w}^{\rm v}$ on the left hand side of Eq.~\eqref{eq:transporttwovalleys}, requires special attention.  Naively interpreted, this term yields generation of net local charge density for $\sum_w\eta_w \EE^{\rm v}_w\neq 0$, which follows from Eq.~\eqref{eq:E5inversion}. However, it has been
shown~\cite{Qi2013anomaly,pikulinfranz2016,Behrends2019anomaly} that this apparent charge non-conservation is unphysical, and pertains to the states within a certain cut-off near the Fermi level. It is compensated by the equal and opposite charge density change in the Fermi sea due to the motion of the band bottom. Therefore, in the present case the charge conservation for the total charge density, $n^{\rm{tot}}$, does not follow from the transport equation itself, but has to be imposed separately:
\begin{align}\label{eq:chargeconservation}
  \partial_t n^{\rm {tot}}=\sum_{w}\partial_t n_w-\sum_w \frac{e^2}{4\pi^2}\eta_w\bm B\cdot\bm{\mathcal{E}}^{\rm v}_w.
\end{align}
This equation describes the fact that the Fermi sea charge does change due to the anomaly, but does not diffuse like the Fermi-surface part of the change density. 

At this point, it is worthwhile to summarize the situation with local charge generation by the anomaly terms in each of the cases of transport equations we have considered, see Eqs.~\eqref{eq:transport}, \eqref{eq:transport_vector}, and~\eqref{eq:transporttwovalleys}. The latter of the three has just been discussed. For vector perturbations in TR-invariant crystals, the second term on the left hand side of Eq.~\eqref{eq:chargeconservation} vanishes, see Section~\ref{sec:vectorTR}. Then it follows from Eqs.~\eqref{eq:transport_vector} that the total charge density changes only due to valley diffusion. Finally, in the scalar case, Eq.~\eqref{eq:transport}, we did allow the anomaly term to generate local charge density. However, in that case, the corresponding charge accumulation rate comes from the divergence of a physical current: the generalized CME current of Eq.~\eqref{eq:CMEcurrent}. This apparent inability of the low-energy transport theory to capture the full physical picture behind various perturbations is a manifestation of the very topology of the Weyl semimetal. The low-energy theory, describing the vicinity of a Weyl point, and being by definition local in momentum space, cannot capture the way the Weyl points are connected through the parts of the band structure away from the low-energy region.~\cite{Haldane2004} 

Turning back to the calculation of the entropy production in centrosymmetric Weyl semimetals, from this point on we can use the considerations of Sec.~\ref{sec:scalarcase}, and impose the charge neutrality condition $n^{\rm{tot}}=0$ at finite frequency as
\begin{align}\label{eq:neutralitytwovalleys}
  \sum_{w} n_w-\frac{i}{\w}\sum_w \frac{e^2}{4\pi^2}\eta_w\bm B\cdot\bm{\mathcal{E}}^{\rm v}_w=0.
\end{align}
Eqs.~\eqref{eq:transporttwovalleys} and \eqref{eq:neutralitytwovalleys} play the role of the transport and charge neutrality equations (Eqs.~\eqref{eq:transport} and \eqref{eq:neutrality}, respectively) already encountered in the case of scalar mechanism. The expression for the entropy production in the case of a scalar mechanism, Eq.~\eqref{eq:entropy_production} is minimally modified:
\begin{align}\label{eq:entropy_production_vector}
    T\dot S=\nu D\sum_{w=\pm}\int_\rr(-\bm\nabla \mu_w+e\bm{\mathcal{E}}^{\rm v}_w)^2+\sum_{w=\pm}\int_\rr\frac{\nu}{\tau_v}\left(\mu_w-\ool\mu\right)^2.
\end{align}

To determine the entropy production in the present case, we recall that the non-equilibrium part of the electrochemical potential in a given valley is related to the density disturbance $n_w$  and the screening potential $e\phi$: $\mu_w=e\phi+n_w/\nu$. Further, it is convenient to introduce $\ool\mu=(\mu_++\mu_-)/2$, $\d\mu=(\mu_+-\mu_-)/2$, and use $\Omega$ and $\tau_q$ defined in Eqs.~\eqref{eq:Omega} and \eqref{eq:tauq}, respectively. Then taking the valley-symmetric and valley-antisymmetric parts of the transport equation~\eqref{eq:transporttwovalleys}, and taking into account the charge neutrality condition~\eqref{eq:neutralitytwovalleys}, we arrive at the following system of equations for the Fourier components of $\ool\mu,\d\mu$, and $e\phi$:
\begin{subequations}
\label{eq:TRtransport}
\begin{align}
  &-i\w(\ool{\mu}-e\phi)+Dq^2\ool\mu=\frac{e^2}{4\pi^2\nu}\bm B\cdot\bm{\mathcal{E}}_5-i\Omega\delta \mu,\label{eq:TRtransport:a}\\
  &(i\w-\frac{1}{\tau_v}-Dq^2)\d\mu=ieD\bm q\cdot\bm{\mathcal{E}}_5+i\Omega \ool\mu,\label{eq:TRtransport:b}\\
  &e\phi=\ool\mu-\frac{ie^2}{4\pi^2\w\nu}\bm B\cdot\bm{\mathcal{E}}_5\label{eq:TRtransport:c}
\end{align}
\end{subequations}
Upon inspecting these equations, it becomes clear that if the electroneutrality condition,  Eq.~\eqref{eq:TRtransport:c}, is used to eliminate $e\phi$, the $\bm B\cdot\bm{\mathcal{E}}_5$ ``anomaly" term disappears from the equations. That is, the motion of the band bottom that it describes gets completely screened out. We can conclude that unless a Weyl semimetal has very few Weyl nodes, and the energies of these points are very close to the Fermi level, the pseudoelectric-field driven anomaly-type physics can manifest itself only in ultrathin films of Weyl semimetals, in which the screening effects are reduced. The estimates presented in Section~\ref{sec:regimes} show that the film thickness should not exceed tens of nanometers. Such small thickness indicates the importance of spatial quantization effects, the investigation of which goes beyond the scope of this paper.

The qualitative picture of the magnetoabsorption in the present model can be inferred from the following considerations. First, we note that a longitudinal $\bm{\mathcal{E}}_5$ field drives an imbalance between valley chemical potentials, since $\qq\cdot\bm{\mathcal{E}}_5\neq 0$ in Eq.~\eqref{eq:TRtransport:b}. The steady value of the chemical potential imbalance is determined by the intervalley scattering and intravalley diffusion. The intravalley electric currents driven by such induced chemical potential gradients are out of phase with the currents driven by $\bm{\mathcal{E}}_5$, hence reduce intravalley Joule heat. This happens regardless of the presence or absence of a magnetic field. The magnetoabsorption for the present model comes from the CME, which couples the equations for $\ool\mu$ and $\d\mu$, via the terms containing $\Omega\propto B$ in Eqs.~\eqref{eq:TRtransport:a} and \eqref{eq:TRtransport:b}, respectively.  The magnetic field affects the entropy production by \emph{reducing} the effective intervalley relaxation time:
\begin{align}\label{eq:tauB}
  \frac{1}{\tau_q}\to \frac{1}{\tau_B}\equiv \frac{1}{\tau_q}+\frac{\Omega^2}{Dq^2},
\end{align}
Hence a magnetic field  \emph{suppresses} the intervalley chemical potential imbalances, and \emph{increases} intravalley currents, leading to positive magnetoabsorption. Note that this is in contrast with the usual negative magnetoresistance mechanism,\cite{Nielsen1983,SonSpivak2013} in which reduction of the intervalley relaxation time would suppress entropy production.

The expression for the entropy production averaged over a period of oscillation can be easily derived from Eqs.~\eqref{eq:TRtransport}. For a longitudinal pseudoelectric field, $\bm{\mathcal{E}}_5=\bm e_{\bm q} \mathcal{E}_5$, we obtain
\begin{align}\label{eq:TRmagnetoabsorption}
  \langle T\dot S^{\rm v}\rangle_{\rm {osc}}=e^2\nu D V\mathcal{E}_5^2\left(1-\frac{Dq^2\tau_B}{1+\omega^2\tau_B^2}\right),
\end{align}
where $\tau_B$ is the effective intervalley relaxation time in the presence of a magnetic field, defined in Eq.~\eqref{eq:tauB}. Using Eq.~\eqref{eq:scalarvsvector} to express $\mathcal{E}_5$ in terms of the vector part of the deformation potential, and restricting ourselves to the limit of $\w,Dq^2\ll\tau_v^{-1}$, and $B\to 0$, we can write the magnetoabsorption of Eq.~\eqref{eq:TRmagnetoabsorption} as 
\begin{align}\label{eq:smallomegavectorTRinv}
  \langle T\d\dot S^{\rm v}(\BB)\rangle_{\rm {osc}}\approx \frac{\nu V}4 \frac{\omega^2D\tau_v}{s^2}\omega^2\tau_v\frac{\omega_c^2}{\mu^2}(\ee_{\BB}\cdot\ee_{\bm q})^2 |\xi u|^2.
\end{align}

We emphasize again that the conclusion about the vector mechanism providing the leading contribution to the electronic part of sound magnetoabsorption in the model of a centrosymmetric Weyl semimetal with just two valleys relied on the complete absence of the scalar contribution in this model. This fact is ensured by strong electronic screening. In general, even in a centrosymmetric Weyl semimetal there is a scalar contribution given by Eq.~\eqref{eq:subleading entropy production}, which is greater than that of Eq.\eqref{eq:smallomegavectorTRinv} by a factor $\tau_v/\tau\sim 10^2$, and has the opposite sign.

\subsection{Symmetry restrictions}\label{sec:symmetry}

The magnitude of the even in B-field part of the magnetoabsorption due to scalar perturbations in TR-invariant Weyl materials crucially depends on whether the chiral anomaly driven by the pseudoelectric fields affects the value of the screening electric potential. The effect of the chiral anomaly on the screening potential is described by Eq.~\eqref{eq:bigphi}. In crystals whose symmetry allows $e\Phi\neq 0$, the magnetoabsorption is described by Eqs.~\eqref{eq:smallomegaScalarB} and is relatively large, while in those with $e\Phi=0$ magnetoabsorption is given by ~\eqref{eq:subleading entropy production}, and is clearly suppressed at low frequencies as compared to that of Eq.~\eqref{eq:smallomegaScalarB}. 

It is clear from Eqs.~\eqref{eq:bigphi} that $e\Phi\propto (\bm e_{\bm q}\cdot\bm e_{\bm B})\chi_{ij}u_{ij}$, where $\chi_{ij}$ is a certain symmetric material tensor, and $u_{ij}$ is the symmetric deformation tensor, as before. Since $\bm e_{\bm q}\cdot\bm e_{\bm B}$ is a pseudoscalar, $\chi_{ij}$ must be a pseudotensor. Therefore, $e\Phi\neq 0$ only in crystals with point groups that allow a symmetric pseudotensor. This symmetry requirement is the same as for nonzero rotatory power in crystals exhibiting the natural optical activity.\cite{LL8,Malgrange2014} That is, a crystal is required to be noncentrosymmetric, and of the 21 noncentrosymmetric point groups, only 15, with the exception of $C_{3h}, D_{3h}, T_d, C_{3v}, C_{4v}, C_{6v}$, will have $e\Phi\neq 0$. (Below, we will refer to these 15 point groups as strongly gyrotropic.)  The leading contribution of the scalar mechanism into magnetoabsorption in the six noncentrosymmetric groups that are not strongly gyrotropic is given by Eq.~\eqref{eq:subleading entropy production}, which exists in any noncentrosymmetric Weyl metal. Furthermore, the contributions of transverse and longitudinal vector perturbations to magnetoabsorption exist in Weyl materials in general, as long as the topological band structure is allowed by symmetry.

We now turn to the acoustic magneto-chiral dichroism. The results for the scalar and longitudinal vector perturbations are given by Eqs.~\eqref{eq:entropyproduction_scalar_odd} and~\eqref{eq:entropyproduction_vector_odd}, respectively.  Acoustic magneto-chiral dichroism is described by the entropy production rate that is odd in the external magnetic field, and the wave vector of the sound wave: $ \langle T\d\dot{S}^{\rm {odd}}(\bm B)\rangle_{\rm osc}\propto (\bm e_{\bm q}\cdot\bm e_{\bm B})\chi_{ijkl}u_{ij}u_{kl}$, where $\chi_{ijkl}$ is a fourth rank pseudotensor, symmetric with respect to the first two indices, the last two indices, and the interchange of the first and last pair of indices. In addition to groups that permit a symmetric second rank pseudotensor, groups $C_{3v}$ and $C_{4v}$ allow a nonzero fourth rank pseudotensor with the above additional symmetries related to permutation of indices. In particular, the case of $C_{4v}$ is special in that this is the point group of the transition metal monopnictide family of Weyl semimetals (see Ref.~\onlinecite{YanFelser2017review} for a review). For $C_{4v}$, $\chi_{ijkl}$ has a single independent component $\chi_{xxxy}=-\chi_{yyyx}$, all other nonzero components can be obtained by appropriate index permutations. Hence in a material with point group $C_{4v}$, the acoustic magneto-chiral dichroism is given by $ \langle T\d\dot{S}^{\rm {odd}}(\bm B)\rangle_{\rm osc}\propto (\bm e_{\bm q}\cdot\bm e_{\bm B})u_{xy}(u_{xx}-u_{yy})\propto (\bm e_{\bm q}\cdot\bm e_{\bm B}) \sin(4\theta) $, where $\theta$ is the angle between the propagation direction and the $a$-axis in the basal plane. It is easy to see that the combination $u_{xy}(u_{xx}-u_{yy})$ is odd with respect to all four mirror operations of $C_{4v}$, as required by the presence of a pseudoscalar $\bm e_{\bm q}\cdot\bm e_{\bm B}$.

To conclude this Section, we point out that the limit of strong electronic screening (electroneutrality limit) employed in this work can impose restrictions beyond those related to a particular point group. For instance, the toy example of a TR-invariant band structure with four Weyl nodes presented at the end of Section~\ref{sec:discussion} below has $C_2$ point group, and should exhibit the acoustic magneto-chiral dichroism. However, this model effectively reduces to a two-node band structure with nodes of opposite chirality that are not related by the TR and two-fold rotation symmetries in the original band structure. For just two independent Weyl points, electroneutrality forbids the acoustic magneto-chiral dichroism by setting the deformation potential averaged over valleys to zero, and nullifying the sum over the valleys on the right hand side of Eqs.~\eqref{eq:entropyproduction_scalar_odd}. However, in Weyl semimetals with a large number of Weyl nodes, like TaAs, the point group symmetry restrictions listed above should be generic.

\section{Discussion of main results}\label{sec:discussion}

In this Section, we compare the magnitudes of various absorption mechanisms discussed in Section~\ref{sec:calculations}. In experiments on sound absorption, one measures the sound attenuation coefficient, $\Gamma(\bm B)$. It is defined as the fraction of acoustic energy dissipated into heat per unit distance traveled by the wave. The attenuation coefficient allows to discuss the results of this paper with most economically written expressions. Assuming losses are small, the attenuation coefficient for various perturbation types can be defined taking the entropy production rates of Section~\ref{sec:calculations} assuming the sound amplitude to be constant throughout the crystal, and dividing them by the energy of the wave, and the speed of sound:
\begin{align}
    \Gamma(\bm B)=\frac{\langle T\d\dot S(\BB)\rangle_{\rm {osc}}}{s U V},
\end{align}
where $U_A=\frac12 \rho s^2 (u_{ij})^2$ is the energy density of the acoustic wave averaged over a period of oscillation, in which $\rho$ is the mass density of the crystal, and $V$ is the volume of the system.

For a particular crystal, its point group determines the dependence of the magnitude of the magnetoabsorption on the propagation direction and polarization of the wave. Below we would like to avoid listing all such details for all relevant point groups, but rather focus on the broad-brush features of the obtained results: the overall magnitude and sign of sound magnetoabsorption for various mechanisms, and its dependence on frequency. To this end, we present our results as dimensionless ratios of the acoustic attenuation coefficients with and without a magnetic field.

First, we summarize our results for TR-invariant Weyl semimetals. We restrict ourselves to the regime in which the dissipation is determined by intervalley scattering. In typical Weyl semimetals, this corresponds to frequencies around a gigahertz, or lower, \textit{i.e.} to frequency intervals $\textbf{\textit{I}}$ and $\textbf{\textit{II}}$ of Fig.~\ref{fig:frequencies}.

\subsection{Scalar perturbations in TR-invariant Weyl semimetals}
We start with the case of scalar perturbations. These produce the largest dissipation rates, unless their contribution is ruled out by symmetry (see Section~\ref{sec:symmetry}), hence can be used as reference to evaluate the magnitude of absorption due to vector perturbations. We list the expressions for the part of the attenuation coefficient that is odd in the magnetic field and describes the acoustic magneto-chiral dichroism, and the part quadratic in magnetic field, which describes magnetoabsorption. 

Since TR-invariant Weyl semimetals are necessarily noncentrosymmetric crystals, the sound attenuation rate, $\Gamma(\bm B)$, in general contains a part odd in the $B$-field, which is also odd in frequency and in the wave vector of the acoustic wave to satisfy the Onsager symmetry. This odd  part of the attenuation rate describes the acoustic magneto-chiral dichroism: sound absorption is different for the opposite propagation directions. This effect exists only in noncentrosymmetric crystals, and is of quantum-mechanical origin. Indeed, the band structure of a TR-invariant crystal still contains a center of inversion.  This implies that a treatment based on the semiclassical Boltzmann equation, which accounts solely for the dispersion of band elections, but not their wave functions, and hence disregards band geometry, will not produce a non-zero magneto-chiral dichroism.  It is also important to note that the so-called ``extrinsic'' effects, related to disorder scattering, cannot compete with the Berry curvature ones in this case, unlike in the familiar one of the anomalous Hall effect.\cite{NagaosaReview} The reason for this is the fact that disorder scattering does not lead to spectral flow between valleys, hence associated perturbations are relaxed on short intravalley time scales. Therefore, the contribution that we obtained from the Berry-curvature related corrections to the semiclassical transport is the leading contribution to the acoustic magneto-chiral dichroism in Weyl semimetals. 

Using Eqs.~\eqref{eq:scalarnoB} and ~\eqref{eq:entropyproduction_scalar_odd}, we obtain the following odd part of the absorption rate for scalar perturbations (superscript `s'), associated with the nodal motion in the energy space:
\begin{align}\label{eq:resultgammaodd}
\frac{\Gamma^{\rm {s}}(\bm B)-\Gamma^{\rm s}(-\bm B)}{\Gamma^{\rm s}(0)}\sim
\mathrm{sgn}(\omega)(\omega\tau_v)^2\frac{v}{s}\frac{\hbar\omega_c}{E_F}(\bm e_{\bm B}\cdot\bm e_{\bm q}).
\end{align}
Here we introduced $E_F$ as the typical doping level counted from the energy of a typical Weyl point in the crystal. 

We can estimate the relative magnitude of the acoustic magneto-chiral dichroism from Eq.~\eqref{eq:entropyproduction_scalar_odd}. As a function of frequency, it reaches its maximum at the boundary between the intervals $\textbf{\textit{II}}$ and $\textbf{\textit{III}}$, $\omega\tau_v\sim \frac{s}{v}\sqrt{\frac{\tau_v}{\tau}}\sim 10^{-2}$, for higher frequencies, intravalley density imbalances are relaxed by very fast diffusion, and the effect diminishes. Furthermore, the applicability of the present theory is limited to classically weak magnetic fields, $\omega_c\tau\lesssim1$. Hence for clean semimetals, $E_F\tau/\hbar\sim 10^2$,  $\omega_c\tau\sim 1$, and , $v/s\sim 10^3$, and $\omega\tau_v\sim 10^{-2}$, one obtains that the acoustic magneto-chiral dichroism is a 0.1\%-effect. The relative magnitude of the acoustic magneto-chiral dichroism goes up by an order of magnitude in moderately disordered Weyl semimetals with $E_F\tau/\hbar\sim 10$, if one can keep a large ratio of $\tau_v/\tau$, which quite typical for existing Weyl semimetals with $E_F\sim 10\mathrm{meV}$, and $\tau\sim 1\mathrm{ps}$. This is a measurable effect. It follows from the discussion in Section~\ref{sec:symmetry} that it can be observed in existing Weyl materials from the transition metal monopnictide family. 

Turning to the even part of the absorption rate, which describes the usual magnetoabsorption, we obtain
\begin{align}\label{eq:resultscalar}
\frac{\Gamma^{\rm {s}}(\bm B)+\Gamma^{\rm s}(-\bm B)}{2\Gamma^{\rm s}(0)}-1\sim -\frac{\tau_v}{\tau}\frac{(\hbar\omega_c)^2}{E_F^2}(\bm e_{\bm B}\cdot\bm e_{\bm q})^2.
\end{align}
The estimate of Eq.~\eqref{eq:resultscalar} was obtained by comparing Eq.~\eqref{eq:smallomegaScalarB} to Eq.~\eqref{eq:scalarnoB}.  The most notable feature of the small-frequency magnetoabsorption of Eqs.~\eqref{eq:smallomegaScalarB}, or Eq.~\eqref{eq:resultscalar} is its negative sign. Physically, it originates from the magnetic field dependence of the screening potential in Weyl semimetals with strongly gyrotropic groups, see Eq.~\eqref{eq:bigphi}, and the discussion of symmetry restrictions in Section~\ref{sec:symmetry} above. As mentioned in Section~\ref{sec:calculations}, the negative sign of the magnetoabsorption stems from the fact that the currents driven by magnetic-field-dependent part of the screening potential act to reduce the intervalley imbalances driven by the deformation potential itself, and hence reduce dissipation.

We note that the relative magnetoabsoprtion in Eq.~\eqref{eq:resultscalar} is independent of frequency for $\omega\to 0$, unlike the odd part of magnetoabsorption, Eq.~\eqref{eq:resultgammaodd}, or the even part of magnetoabsorption in Weyl semimetals with weakly gyrotropic point groups, see Eq.~\eqref{eq:resultscalarweak} below. 

In crystals with point groups $C_{3h}$, $D_{3h}$, $T_d$, $C_{3v}$, $C_{4v}$, $C_{6v}$, a magnetic-field-dependent part of the screening potential is absent (Section~\ref{sec:symmetry}). In this case, one obtains from Eqs.~\eqref{eq:scalarnoB} and~\eqref{eq:subleading entropy production} that
\begin{align}\label{eq:resultscalarweak}
\frac{\Gamma^{\rm {s}}(\bm B)+\Gamma^{\rm s}(-\bm B)}{2\Gamma^{\rm s}(0)}-1\sim -\frac{v^2}{s^2}(\omega\tau_v)^2\frac{(\hbar\omega_c)^2}{E_F^2}(\bm e_{\bm B}\cdot\bm e_{\bm q})^2.
\end{align}
As compared with sound absorption in strongly gyrotropic groups, Eq.~\eqref{eq:resultscalar}, the result of Eq.\eqref{eq:resultscalarweak} is obviously suppressed at low frequencies, but the two become comparable for $\omega\tau\sim \frac{s}{v}\sqrt{\frac{\tau}{\tau_v}}$, which is the crossover region between intervals $\textbf{\textit{II}}$ and $\textbf{\textit{III}}$ of Fig.~\ref{fig:frequencies}.

We can also compare Eq.~\eqref{eq:resultscalar} to the classical magnetoabsorption that stems from the cyclotron motion of carriers. Typically, such magnetoabsorption is negative, and its relative magnitude is set by $(\omega_c\tau)^2$. It then follows from Eq.~\eqref{eq:resultscalar} that the relative magnitude of the classical and anomaly-related mechanisms of the sound magnetoabsorption in strongly gyrotropic crystals is controlled by $\frac{\tau_v}{\tau}\frac{\hbar^2}{(E_F\tau)^2}$. For $\tau_v/\tau\sim 10^{2}$, $E_F\tau/\hbar\sim 10^2$, this parameter is of order of $10^{-2}$. Hence the anomaly-related contribution to magnetoabsorption of sound is small in clean Weyl semimetals. However, the aforementioned parameter can become of order of unity in moderately disordered materials. Indeed, for $E_F\sim 10\,\mathrm{meV}$ and $\tau\sim 1\,\mathrm{ps}$, we obtain $E_F\tau/\hbar\sim 10$, and $\frac{\tau_v}{\tau}\frac{\hbar^2}{(E_F\tau)^2}\sim 1$. While the overall magnitudes and signs of the two contributions are the same in this case, the anomaly-related one can still be detected using its angular dependence: there should be a substantial decrease in magnetoabsorption for sound propagating along the magnetic field. In the case of crystals with weakly noncentrosymmetric groups, the anomaly-related sound magnetoabsorption is negative, is suppressed at low frequencies by an additional factor of $(\omega\tau_v)^2$, see Eq.~\eqref{eq:resultscalarweak}. However, it can also of magnitude comparable to the classical contribution as finite frequencies. For a moderately disordered Weyl semimetal, this should happen for $\omega\tau_v\gtrsim \frac{E_F\tau}{\hbar}\frac{s}{v}\sim 10^{-2}$, or $\omega\gtrsim 10^8\,\mathrm{s^{-1}}$.

\subsection{Vector perturbations in TR-invariant Weyl semimetals}
We now discuss sound absorption due to transverse and longitudinal vector perturbations in TR-invariant Weyl semimetals. These correspond to Weyl nodes motion in the Brillouin zone, see Fig. ~\ref{fig:conemotion}.  To enable easy comparison with the scalar case, we will normalize the attenuation rates for the vector perturbations by the corresponding ones for the scalar case.

The sound absorption due to vector perturbations is in general subdominant as compared to the scalar ones, with some inconsequential exceptions that will be discussed below. To illustrate this point, we first compare the attenuation rate due to the transverse vector perturbations in the absence of a magnetic field to the one coming from the scalar mechanism. From Eqs.~\eqref{eq:scalarnoB} and~\eqref{eq:vectortransversenoB}, we can estimate the ratio of the two attenuation coefficients as
\begin{align}
    \frac{\Gamma^{\rm {v}}(0)}{\Gamma^{\rm {s}}(0)}\sim
    \begin{cases}
    \frac{\tau}{\tau_v}, \textrm{intervals \textit{I, II}},\\
    \frac{s^2}{v^2}\left(\frac{Dq^2}{\omega}\right)^2, \textrm{interval \textit{III}},
    \end{cases}
\end{align}
where the frequency intervals \textit{I-III} are those pertaining to Fig.~\ref{fig:frequencies}, and we assumed similar magnitudes of scalar and vector parts of the deformation potential.\cite{magnitudes_of_deformpotentials} The diffusive regime is bounded by the condition $Dq^2\lesssim 1/\tau$, or $\omega\tau\lesssim s/v$. Hence in the absence of a magnetic field the entropy production due to the transverse vector  mechanism is small as compared to the scalar mechanism, but the two can become comparable at the edge of validity of the present approach. The same conclusion holds for the longitudinal vector perturbations, see Eq.~\eqref{eq:vectorlongitudinalnoB}.

Turning to sound absorption in magnetic field for vector perturbations, we note that only longitudinal vector perturbations lead to the acoustic magneto-chiral dichroism. We obtain from Eqs.~\eqref{eq:entropyproduction_vector_odd} and~\eqref{eq:entropyproduction_scalar_odd} that the relative magnitude of the acoustic magneto-chiral dichroism for vector and scalar cases is
\begin{align}\label{eq:resutsvectorodd}
    \frac{\Gamma^{\rm {v,l}}(\bm B)-\Gamma^{\rm v,l}(-\bm B)}{\Gamma^{\rm {s}}(\bm B)-\Gamma^{\rm s}(-\bm B)}\sim\frac{\tau}{\tau_v}\ll1. 
\end{align}
Given that the symmetry restrictions for the acoustic magneto-chiral dichroism are the same for both types of perturbations, we see that the acoustic magneto-chiral dichroism due to the vector perturbations is always subdominant as compared to the one for scalar perturbations. 

In TR-invariant Weyl semimetals, the sound magnetoabsorption due to vector perturbations is positive, its sign being of the same origin as the positive transport magnetoconductance in Weyl semimetals. The two problems are similar since vector perturbations are not subject to screening in TR-invariant Weyl semimetals. However, in crystals with strongly gyrotropic groups, such that Eq.~\eqref{eq:resultscalar} holds, the magnetoabsorption due to both longitudinal and transverse perturbations is always small as compared to the scalar case. This can be seen from Eqs.~\eqref{eq:smallomegaScalarB}, \eqref{eq:vtB}, and \eqref{eq:smallomegavectorB}, which yield 
\begin{align}\label{eq:resultsvectorB}
\frac{\left|\Gamma^{\rm {v}}(\bm B)+\Gamma^{\rm v}(-\bm B)\right|}{\Gamma^{\rm {s}}(0)}\sim\frac{(\hbar\omega_c)^2}{E_F^2}(\bm e_{\bm B}\cdot\bm e_{\bm q})^2. 
\end{align}

It is then clear that the magnetoabsorption for vector  perturbations, regardless of their longitudinal or transverse nature, satisfies
\begin{align}\label{eq:resultsvectorBrelative}
          \left|\frac{\Gamma^{\rm {v}}(\bm B)+\Gamma^{\rm v}(-\bm B)-2\Gamma^{\rm {v}}(0)}
          {\Gamma^{\rm {s}}(\bm B)+\Gamma^{\rm s}(-\bm B)-2\Gamma^{\rm {s}}(0)}\right|\sim\frac{\tau}{\tau_v}\ll1. 
\end{align}
The weakness of magnetoabsorption due to vector perturbations in TR-invariant Weyl semimetals with strongly gyrotropic groups makes its sign not relevant experimentally. 

In crystals with point groups that forbid the leading term in the entropy production due to scalar perturbation, Eq.~\eqref{eq:resultscalar}, the vector perturbations do provide the leading anomaly-related contribution to the sound magnetoabsorption at low frequency. This follows from the leading nonzero scalar mechanism's contribution in such crystals, now given by Eq.~\eqref{eq:resultscalarweak}, being suppressed by an additional factor of $\omega^2$ as compared to the vector contribution, Eq.~\eqref{eq:resultsvectorB}. However, one can easily show that the positive magnetoabsorption due to vector contributions is always small as compared to the negative classical magnetoabsorption by a factor of $1/E_F^2\tau^2\ll1$. We thus conclude that vector perturbations are not important for sound absorption in TR-invariant Weyl semimetals: scalar perturbations always dominate sound absorption in zero field; the acoustic magneto-chiral dichroism due to longitudinal vector perturbations is always small as compared to the one due to scalar perturbations; magnetoabsorption due to vector perturbations is small compared to that of scalar one in crystals with strongly gyrotropic groups, and is small compared to the classical magnetoabsorption in other noncentrosymmetric crystals. 

\subsection{Vector perturbations in centrosymmetric Weyl semimetals}
We now turn to the case of centrosymmetric Weyl semimetals with broken TR symmetry, focusing on a toy two-valley model. In such a model, the scalar mechanism contribution is fully suppressed. The reason for that is the fact that the inversion symmetry dictates the deformation potentials in the two valleys be the same. This means that they act as a conventional electric potential, and are fully screened out.

For centrosymmetric Weyl semimetals with broken TR symmetry, the magneto-absorption is positive (see the discussion around Eq.~\eqref{eq:tauB}), and is an even function of the magnetic field. Using Eqs.~\eqref{eq:TRmagnetoabsorption}, and~\eqref{eq:smallomegavectorTRinv}, we obtain for the relative magnetoabsorption:
\begin{align}\label{eq:resultsvector}
\frac{\Gamma^{\rm v}(B)-\Gamma(0)}{\Gamma(0)}\sim (\omega\tau_v)^2\frac{v^2}{s^2}\frac{(\hbar\omega_c)^2}{E_F^2}(\bm e_{\bm B}\cdot\bm e_{\bm q})^2.
\end{align}

Since the usual magneto-absorption due to the suppression of Joule heating by cyclotron motion is negative and is not suppressed at low frequencies, the net magneto-absorption is always negative for $\omega\to 0$ in centrosymmetric Weyl semimetals. However, one can expect a sign reversal of the effect at finite $\hbar\omega\sim \frac{s}{v}\frac{\tau}{\tau_v}E_F\sim 10^{-5}E_F$ due to the anomaly related contribution growing with frequency. For $E_F\sim 10\mathrm{meV}$, the crossover is expected at $\omega\sim 10^8\mathrm{s}^{-1}$.

To finish the discussion of the centrosymmetric case, we observe that in Weyl semimetals with more than two nodes, in which the restrictions imposed by charge neutrality are not as severe, the scalar perturbation magnetoabsorption of Eq.~\eqref{eq:resultscalar} is still absent due to assumed inversion symmetry of the crystal. However, even the subleading contribution to magnetoabsorption, Eq.~\eqref{eq:resultscalarweak}, dominates over the one related to vector perturbation by a factor of $\tau_v/\tau\gg1$, which is evident from Eq.~\eqref{eq:resultscalarweak}, and Eq.~\eqref{eq:resultsvector} given below. Therefore, we expect the change in the sign of the magnetoabsorption only in a material with very few, ideally two, nodes. This is the situation expected in $\mathrm{EuCd}_2\mathrm{As}_2$.\cite{Ma2019EuCdAs,Soh2019EuCdAs}

\subsection{Sound absorption in a minimal model of a Weyl semimetal}\label{sec:simplemodel}
Finally, to illustrate the general frequency and magnetic field dependence of sound absorption given by Eq.~\eqref{eq:scalartotal} in the least cumbersome way, we apply it to a minimal model of a TR-invariant Weyl metal, depicted in Fig.~\ref{fig:weylpoints}a. This model consists of two pairs of Weyl nodes, the nodes within each pair being connected by the time-reversal symmetry and thus having the same chirality, opposite for each pair. Given the simplifying assumptions under which Eq.~\eqref{eq:scalartotal} was obtained, it is appropriate to assume that the deformation potential within each nodes is proportional to a unit matrix, $\lambda_{w,ab}=\lambda_w\d_{ab}$, such that for nodes with positive chirality ($\eta_w=+1$) one has $\lambda_w\equiv \lambda_+$, and for the nodes with the negative chirality ($\eta_w=-1$), one has $\lambda_w\equiv \lambda_-$. Such configuration of the deformation potentials supplies the model with a notion of chirality, and is appropriate for a Weyl semimetal with point group $C_2$, which is a strongly gyrotropic group.

For this simple TR-invariant model, the magneto-chiral dichroism vanishes. This is an artifact of the $N_v=4$ model, in which the deformation potentials of the different TR-related pairs of nodes are equally spaced from the average potential, but have opposite chiralities, forcing them to cancel each other's contributions. This can be seen by substituting the deformations potentials of the minimal model into Eq.~\eqref{eq:entropyproduction_scalar_odd}.

\begin{figure*}
     \centering
  \includegraphics[width=7.0in]{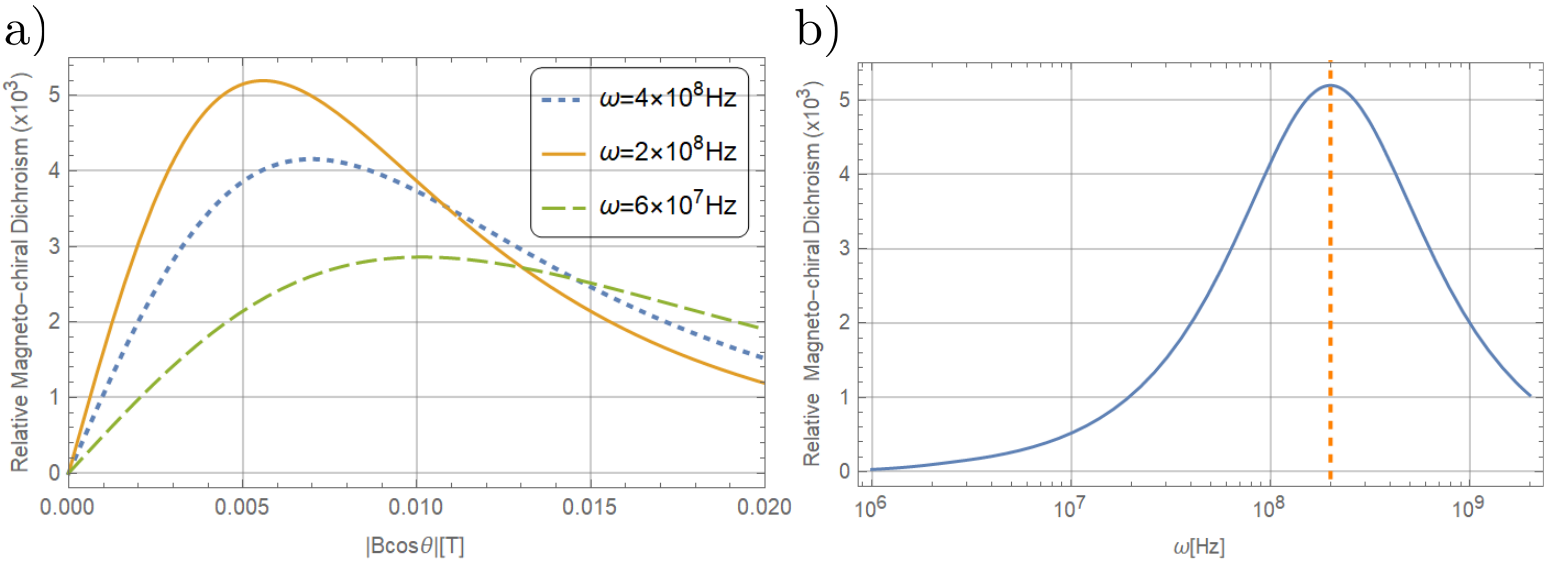}
    \caption{Relative magnitude of the acoustic magneto-chiral dichroism as defined in Eq.~\ref{eq:resultgammaodd}, for a TR invariant Weyl semimetal. Scalar deformations for each nodal pair are chosen to be $\left.\lbrace\lambda,2\lambda,3\lambda,4\lambda\rbrace\right.$ with respective chiralities $\left.\lbrace+1,-1,-1,+1\rbrace\right.$ which provide a strongly gyrotropic model. Parameters are the same as in Fig.~\ref{fig:scalar_zero_field} along with $\mu=10meV$. For the magnetic field plot (a) the horizontal axis is $\bm B\cdot  \bm e_{\bm q} \propto B\cos{\theta}$ and terminates at $\omega_c\tau=1$, as in Fig.~\ref{fig:relative_magnetoabsorption}. The frequency plot (b) corresponds to the peak of the magnetic field response. The horizontal axis terminates at $\omega\tau=1$, with the dashed line highlighting the crossover between regimes ($\textbf{\textit{I}}-\textbf{\textit{II}}$) and ($\textbf{\textit{III}}$) of Fig.~\ref{fig:frequencies}, identically to the dashed line in Fig.~\ref{fig:scalar_zero_field}. This crossover frequency can be seen to maximize the magneto-chiral dichroism at optimal field strength. Note the magnitude of the effect is bounded by $~0.5\%$.}\label{fig:amcd}
\end{figure*}

For a general TR-invariant material with $N_v>4$, we expect  Eq.~\eqref{eq:entropyproduction_scalar_odd} to produce a odd-in-$B$ magnetoabsorption. Using expressions obtained in this work, in particular Eq.~\eqref{eq:scalartotal}, we can still outline the frequency and magnetic field dependence of acoustic magneto-chiral dichroism in a generic noncentrosymmetric crystal with a point group that allows this effect. Restricting ourselves only to illustrate the general features of B-field and frequency dependencies of the magneto-chiral effect, we choose a hypothetical model that consists of $N_v=8$ nodes, and set the deformation potentials of the four TR-related pairs to be $\lbrace \lambda,2\lambda,3\lambda,4\lambda\rbrace$, corresponding to chiralities of $\lbrace +1,-1,-1,+1\rbrace$, respectively. The resulting frequency and magnetic field dependencies of the acoustic magneto-chiral dichroism are illustrated in Fig.~\ref{fig:amcd}.

\begin{figure}
  \centering
  \includegraphics[width=3.in]{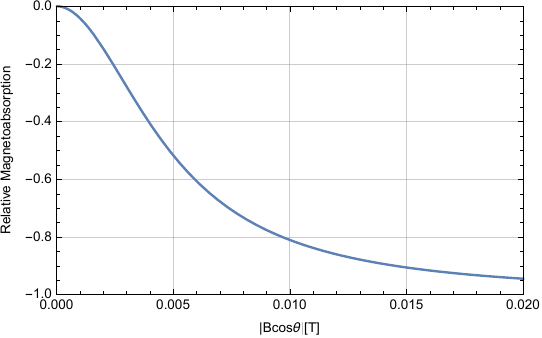}
  \caption{
  Relative magnetoabsorption for a minimal model of a TR invariant Weyl semimetal in the intervalley dominated regime. The horizontal axis combines both the field magnitude and orientation, as the magnetoabsorption is a function of $\Omega\propto\bm B\cdot  \bm e_{\bm q} \propto B\cos{\theta}$.   Parameters are the same as in Fig.~\ref{fig:scalar_zero_field} along with $\mu=10meV$.
  The magnitude of the magnetic field is limited by its being classically weak, $\omega_{c}\tau\leq1$. 
  }\label{fig:relative_magnetoabsorption}
\end{figure}

The minimal $N_v=4$ model does produce a change in even-in-magnetic field part of sound magnetoabsorption, which is quadratic for small fields. The result is presented in Eq.~\eqref{eq:resultscalar}. Aside from its surprising negative sign, discussed in the following Section, the relative absorption is frequency independent, as opposed to Eq.~\eqref{eq:resultgammaodd} which is negligible at low enough frequencies. Higher orders in magnetic field can be obtained from the full solution of Eq.~\eqref{eq:scalartotal}, which yields a Lorentzian-type function of $\Omega/\omega$ that remains frequency independent within the intervalley dominated regime. The relative magnetoabsorption for this minimal model is depicted in Fig.~\ref{fig:relative_magnetoabsorption}.

\subsection{Relation to previous works and physical interpretation of magnetoabsorption due to scalar perturbations}\label{sec:previouswork}

To conclude this Section, we would like to summarize the relationship between the results for scalar perturbations obtained in this paper and previous works, as well as provide a physical interpretation of our main results. 

The existence of an odd-in-$\BB$ part of the entropy production was noticed in Ref.~\onlinecite{Garate2020phonon}. However, the expression obtained in our work, Eq.~\eqref{eq:entropyproduction_scalar_odd}, contains an additional factor of $\omega^2\tau_v^2$ as compared to its counterpart from Ref.~\onlinecite{Garate2020phonon}, and has a different structure with respect to how the valley-specific deformation potentials enter it. In particular, the expressions obtained in our work are insensitive to a valley-independent shift in the deformation potentials, $\lambda_w\to\lambda_w+\delta \lambda$. It is not the case for Ref.~\onlinecite{Garate2020phonon}, despite the fact that it does take into account strong screening. The two sets of results on the magneto-chiral dichroism can be brought in correspondence~\cite{GarateInprivate} by correcting the computational procedure employed in Ref.~\onlinecite{Garate2020phonon}.

Anomaly-related contribution to sound absorption in the semiclassical regime was considered in Ref.~\onlinecite{andreevspivak2016}. In that work, magneto-chiral dichroism was not considered. It was also implicitly assumed that the screening potential was independent of the magnetic field, which is equivalent to setting  $e\Phi=0$ in the context of our work. Therefore, the present work's results for the acoustic magneto-chiral dichroism, Eq.~\eqref{eq:resultgammaodd}, and the even-in-B part of magnetoabsorption in strongly gyrotropic crystals, Eq.~\eqref{eq:resultscalar}, could not be obtained within the framework of Ref.~\onlinecite{andreevspivak2016}. However, even the results of Eqs.~\eqref{eq:subleading entropy production}, or~\eqref{eq:resultscalarweak}, obtained for the situation in which $e\Phi=0$, are of opposite sign as compared to their counterpart from Ref.~\onlinecite{andreevspivak2016}, and also contain an additional factor of $\omega^2\tau_v^2$. 

The observation of the sign of the even-in-B part of the magnetoabsorption being negative, made here, is somewhat counter-intuitive. At least in the $\omega\to 0$ limit, which corresponds to interval $\textbf{\textit {I}}$ of Fig.~\ref{fig:frequencies}, and in which diffusive currents are small, one naively expects the usual DC transport theory\cite{SonSpivak2013} to determine the electronic response, which typically gives \textit{positive} magnetoconductance due to the chiral anomaly~\cite{Nielsen1983, SonSpivak2013} in Weyl metals. Given this breakdown of the naive relationship between the magnetotransport and sound magnetoabsorption problems, as well as the discrepancies we listed above, it appears worthwhile to develop a simple physical picture of sound magnetoabsorption in Weyl semimetals, which is presented below.

In Weyl semimetals, the key difference between the magnetotransport and sound magnetoabsorption problems is the fact that sound wave propagation in general leads to valley electrochemical potential imbalances even \emph{without} an external magnetic field. Being quadratic in the sound wave amplitude, sound magnetoabsorption then comes from an interplay of electrochemical imbalances created with and without a B-field. Below we elaborate on this interplay for frequencies at which the sound absorption is dominated by intervalley scattering, and weak magnetic fields. 
In what follows, we focus on the frequency interval $\textbf{\textit{I}}$, 
see Fig.~\ref{fig:frequencies}, in which $1/\tau_v\gg \w\gg Dq^2$, and sound absorption is dominated by intervalley scattering. 

In the absence of a magnetic field, and in the electroneutrality limit associated with strong electronic screening, see Eq.~\eqref{eq:neutrality}, it follows from Eq.~\eqref{eq:elpot} that the screening potential is given by $e\phi(B=0)=-\ool{\lambda}u$. This result has a simple interpretation: the screening potential makes sure there is no average force acting on the conducting electrons, which means there is no net longitudinal current in the system. This implies that no charge accumulation is generated, as required by charge neutrality. In turn, summing Eq.~\eqref{eq:elchempot} over the valley index, and using $e\phi(B=0)=-\ool{\lambda}u$, one can readily see that for $B=0$ the electrochemical potential averaged over all valleys satisfies $\ool \mu=0$. It then follows from Eq.~\eqref{eq:chempot} that in the limit $1/\tau_v\gg \w\gg Dq^2$, the nonequilibrium part of the electrochemical potential in valley $w$ is given by  
\begin{align}\label{eq:chempotdiscussion}
    \mu_w^{(0)}=-i\omega \tau_\omega(\lambda_w-\ool\lambda)u,\,\,\tau_\omega\equiv\frac{\tau_v}{1-i\omega\tau_v}.
\end{align}
The appearance of the factor $-i\omega\tau_\omega$ in the above expression comes from the fact that at $\omega\to 0$ intervalley scattering quickly equalizes electrochemical potentials of all valleys. To zeroth order in $\omega\tau_v$, this results in $\mu_w=0$ even if $(\lambda_w-\ool\lambda)\neq 0$, and Eq.~\eqref{eq:chempotdiscussion} follows. It is the electrochemical potential imbalances of Eq.~\eqref{eq:chempotdiscussion} that leads to entropy production in the absence of a B-field, see Eq.~\eqref{eq:scalarnoB}. 

Having understood the mechanism behind entropy production at zero-field, we move on to $B\neq 0$, and treat the magnetic field perturbatively. In the presence of a B-field, valley electrochemical potential imbalances given by Eq.~\eqref{eq:chempotdiscussion} can lead to a nonzero CME current, see Eq.~\eqref{eq:CMEcurrent} which flows in the direction of the B-field. If the sound propagation direction is also along the B-field, the electrochemical potential imbalance has a space dependence that which leads to the a divergence of the CME current, and as a result, to a B-dependent correction to valley electrochemical potentials. By iterating this argument, one can calculate the sound magnetoabsorption to any order in the B-field. 

To illustrate the preceding argument, let us calculate the correction to the screening potential induced by the magnetic field, which is denoted with $\Phi$ in Eqs.~\eqref{eq:elpot} and~\eqref{eq:bigphi}. If the CME current has a nonzero divergence, there must be a correction to the screening potential that drives the usual Ohmic current with the opposite sign of divergence, to preserve charge neutrality. This correction is thus obtained from equating the divergence of the Ohmic current it drives, $e^2 \nu D N_v q^2 \Phi$, to the negative of the divergence of the generalized CME current of Eq.~\eqref{eq:CMEcurrent}, $-\frac{e^2}{4\pi^2}(i\bm q\cdot \bm B)\sum_{w}\eta_w\mu_w$, where $\mu_w$'s are given by Eq.~\eqref{eq:chempotdiscussion}. As a result, we obtain 
\begin{align}\label{eq:bigphidiscussion}
    e\Phi=-\frac{\omega\Omega}{Dq^2}\ool{\eta\lambda}u,
\end{align}
which is the small-$\omega$, small-B limit of Eq.~\eqref{eq:bigphi}. As before, $\Omega=\frac{e^2}{4\pi^2\nu}\bm q\cdot\bm B$, and we used $\sum_w\eta_w\ool\lambda u=0$ to write this equation. Note that if $q$ and $\omega$ are related by the sound dispersion relation, $\omega =sq$, $e\Phi$ in Eq.~\eqref{eq:bigphidiscussion} is independent of frequency. The corresponding electric field, however, does vanish in the $\omega\to 0$ limit. 

The appearance of a screening potential linear in the magnetic field, Eq.~\eqref{eq:bigphidiscussion}, is already enough to understand the negative sign of the small-B magnetoabsorption given by Eq.~\eqref{eq:resultscalar}. The electric field that is generated by the gradient of $e\Phi$ drives the chiral anomaly, if it has a nonzero projection onto the B-field, \textit{i.e.} when $\bm q\cdot\bm B\neq0$. This generates a quadratic in the B-field correction to valley electrochemical potentials, the sign of which is such as to reduce the electrochemical potential differences that existed without the magnetic field, and gave rise to $e\Phi$. This reduces the dissipation due to the intervalley scattering, hence the negative sign in the right hand side of Eq.~\eqref{eq:resultscalar}.

To see what happens in more detail, and be able to understand the magnetoabsorption regardless of $e\Phi$ being zero or not, as well as the exact origin of the difference between the transport and sound propagation problems, we need to outline the calculation of valley electrochemical potentials, $\mu_w-\ool\mu$, to quadratic order in the B-field. As before, we restrict ourselves to the $1/\tau_v\gg\omega\gg Dq^2$ limit. The steps of this procedure are as follows: we first calculate the generalized CME current in  Eq.~\eqref{eq:CMEcurrent} using the $B=0$ values of $\mu_w$'s from Eq.~\eqref{eq:chempotdiscussion}; then we calculate corrections to the electrochemical potentials due to the valley charge accumulation to the divergence of these currents, balancing it against intervalley scattering at finite frequency (keeping the time derivative terms in the transport equation); after that, we repeat this procedure once again, this time using the linear-in-B corrections to the electrochemical potentials, obtaining the quadratic corrections. We will not present the details of this simple calculation. We only note that that $\ool \mu=e\Phi$ in the electroneutrality limit, which follows from Eqs.~\eqref{eq:elchempot},~\eqref{eq:neutrality}, and~\eqref{eq:elpot}, and put down the answer for the electrochemical potentials to second order in the magnetic field: 
\begin{align}\label{eq:chempotcorrection}
    \mu_w-\ool\mu= \mu_w^{(0)}+\delta\mu_w^{(1)}+\delta\mu_w^{(2)},
\end{align}
where $\mu_w^{(0)}$ is given by Eq.~\eqref{eq:chempotdiscussion}, while $\delta\mu_w^{(1,2)}$ are of order $O(B^{1,2})$, respectively, and are given by 
\begin{align}\label{eq:deltamu12}
    \mu_w^{(1)}&= -i\eta_w\Omega\tau_\omega\mu_w^{(0)},\nonumber\\
    \mu_w^{(2)}&=-\Omega^2\tau_w^2\mu_w^{(0)}-i\eta_w\Omega\tau_\omega e\Phi. 
\end{align}

Eqs.~\eqref{eq:chempotcorrection} and~\eqref{eq:deltamu12} allow to understand the origin of the acoustic magneto-chiral dichroism, and the negative sign of the sound magnetoabsorption. Both are related to $\mu_w^{(0)}\neq0$ in the sound absorption problem. 

First, let us discuss the opposite signs of the even-in-B part of the anomaly-induced sound magnetoabsorption and magnetoconductance. Of the three terms on the right hand side of Eq.~\eqref{eq:chempotcorrection}, $\mu_w^{(1)}$  is the electrochemical potential correction that is analogous to the one associated with the positive magnetoconductance in the transport problem.~\cite{SonSpivak2013} The correspondence is that $i\bm q\mu_w^{(0)}$ be replaced according to $i\bm q\mu_w^{(0)}\to -eE_{\mathrm{tr}}$, where $E_{\mathrm{tr}}$ is the transport electric field. Crucially, in the $\omega\to0$ limit, when $\tau_\omega\to \tau_v$, this `transport' correction as a complex number is shifted by a phase of $\pi/2$ with respect to the other two terms on the right hand side of Eq.~\eqref{eq:chempotcorrection}. Physically, this phase shift is due $\mu_w^{(1)}$ being linear in the gradient of the electrochemical potential unperturbed by the magnetic field. At the same time, in the transport problem valley electrochemical imbalances appear only due to the magnetic field, hence the $O(B^0)$ term, $\mu_w^{(0)}$, is absent in Eq.~\eqref{eq:chempotcorrection} in the transport problem. Since magnetoresistance is determined by $|\mu_w-\ool\mu|^2$, in the transport problem $\delta \mu_w^{(2)}$ can be neglected at small B-fields, as it leads to $O(B^4)$ corrections. 

Contrary to the transport problem, in the sound absorption problem $\delta \mu_w^{(2)}$ cannot be neglected, since $\mu_w^{(0)}\delta \mu_w^{(2)}\sim O(B^2)$. In fact, this term completely cancels the positive `transport' part of the even-in-B magnetoabsorption, and leads to the negative magnetoabsorption of Eq.~\eqref{eq:resultscalar} when $e\Phi\neq 0$, or Eq.~\eqref{eq:resultscalarweak} when $e\Phi=0$.

Turning to the magneto-chiral dichroism, we note that it stems from the fact that at finite frequencies, $\tau_\w$ is not a real number, $\tau_\w\approx \tau_v+i\omega\tau_v^2$, and thus the phase shift between $\mu_w^{(0)}$ and $\delta\mu_{w}^{(1)}$ is no longer $\pi/2$. Because of this, there is an $O(B)$ term appearing in $|\mu_w-\ool\mu|^2$, which is the origin of the acoustic magneto-chiral dichroism. It can be easily shown that this term leads to Eq.~\eqref{eq:entropyproduction_scalar_odd}.

\section{Conclusions}\label{sec:conclusions}

In this work, we studied the electronic contribution to sound magnetoabsorption in semimetallic multivalley systems. Our aim was to highlight the role of non-trivial momentum space topology, namely that of band crossings resulting in isolated Weyl points, functioning as Berry curvature monopoles. 

The key outcome of this work is that the existence of Berry monopoles in the band structure of a semimetal is best inferred from the acoustic magneto-chiral effect. It is a rather small effect, the relative strength of which we estimate to be at $0.1-1\%$ level at optimal frequencies. In practice, these correspond to $\omega\sim 10^{8}\,\mathrm{s^{-1}}$.  However, it is a measurable effect using modern experimental techniques. Symmetry-wise, the acoustic magnetochiral effect can be observed in crystals with 15 strongly gyrotropic groups (which also allow natural optical activity), and crystals with groups $C_{3v}$, and $C_{4v}$. That is, it should be possible to measure it in available Weyl semimetals from the transition metal monopnictide family, \textit{e.g.} TaAs.

Unlike the positive magnetoconductance in transport measurements, the part of the sound magnetoabsorption that is even in the magnetic field is typically negative in TR-invariant Weyl semimetals. Being negative in sign, it can be detected experimentally via its angular dependence, since it is sensitive to the angle between the propagation direction and the magnetic field. In crystals with strongly gyrotropic groups, this additional anomaly-related magnetoabsorption is present at all frequencies. Otherwise, it is significant only at high enough frequencies, $\omega\gtrsim 10^{8}\mathrm{s^{-1}}$. A special case is represented by centrosymmetric Weyl semimetals with broken TR symmetry, and very few Weyl points. In such materials, sound magnetoabsorption is classical and negative at low frequencies, but can change sign as sound frequency is increased due to an anomaly-type contribution. This situation should be relevant for $\mathrm{EuCd_2As_2}$.

\acknowledgements
The work of DAP was supported by the National Science Foundation Grant No. DMR-1853048; the work of OA and RI was supported by the Israel Science Foundation Grant No. 1790/18; the work of AVA was supported by the U.S. Department of Energy Office of Science, Basic Energy Sciences under Award No. DE-FG02-07ER46452 and by the National Science Foundation Grant MRSEC DMR-1719797. We thank Ion Garate, Dmitry Pikulin, and Boris Spivak for useful discussions.

\textit{Note added:} After the submission of this manuscript, another preprint appeared, Ref.~\onlinecite{sukhachov2021anomalous}, which treated the problem of sound mangetoabsorption in Weyl materials due to scalar perturbations. The results of Ref.~\onlinecite{sukhachov2021anomalous} are in agreement with the ones of this work that pertain to the case of the scalar deformation potential.

\appendix 

\section{Derivation of the entropy production and macroscopic transport equations}\label{sec:entropy_appendix}
Considerations of Section~\ref{sec:scalarcase} show that in most frequency regimes the scalar mechanism associated with nodal point motion in the energy space due to the momentum-independent part of the deformation potential is the leading source of sound absorption in Weyl metals. Therefore, in this Appendix we present a derivation of transport equation, Eq.~\eqref{eq:transport}, and the equation for the entropy production, Eq.~\eqref{eq:entropy_production}, from which the results of Section~\ref{sec:scalarcase} were obtained. 
We work in the system of units with $\hbar=k_B=1$. 

For readability, we briefly repeat various definitions from the main text.  We will assume that the temperature is low enough such that near each Weyl point only one of the corresponding conduction or valence bands has a Fermi surface. (It does not matter which band it is.) In what follows we will enumerate both the Weyl points and the Fermi surfaces around them with index $w$. The scalar perturbation is defined as the motion of a Weyl node in energy space due to an acoustic perturbation, such that the unperturbed energy of the band with a Fermi surface, $\varepsilon_w(\pp)$, is transformed into
\begin{align}\label{appendix:totalenergy}
    E_w(\pp)=\varepsilon_w(\pp)+\lambda_wu+e\phi. 
\end{align}
In this expression $\lambda_wu\equiv\lambda_{w,ij}u_{ij}$ is the Weyl point's energy shift associated with the deformation potential $\lambda_{w,ij}$, $u_{ij}$ is the deformation tensor, and $\phi$ is the screening electric potential.

Kinetics of the electrons near a Weyl point/Fermi surface with index $w$ is described by the Boltzmann kinetic equation for their distribution function, $f_w(\pp)$. The kinetic equation with the semiclassical corrections associated with the Berry curvature is given by\cite{SonSpivak2013,MaPesin2015}
\begin{widetext}
\begin{align}\label{appendix:kinetic equation}
    \p_t f_{w}(\pp)+\frac{1}{D_{\bm B}}\left(\bm v_w-e\bm B(\bm v_w\cdot\bm{\mathcal{F}}_w)\right)\bm \nabla_\rr f_w
    +\frac{1}{D_{\bm B}}\left(e\bm {\mathcal{E}}_w+e\bm v_w\times\bm B-e^2(\bm {\mathcal{E}}_w\cdot\bm B)\bm{\mathcal{F}}_w\right)\bm \nabla_\pp f_w=I_w^{\rm {intra}}+I_w^{\rm {inter}},
\end{align}
\end{widetext}
where $v_\pp=\bm\nabla_\pp\varepsilon_w(\pp)$, $D_{\bm B}=1-e\bm B\cdot\bm {\mathcal{F}}_w$, and $e\bm{\mathcal{E}}_w=-\bm\nabla_\rr(\lambda_wu+e\phi)$ in the effective electric field acting on the electrons in valley $w$. Collision integrals $I_w^{\rm {intra}}$, $I_w^{\rm {inter}}$, which describe the intra- and intervalley scattering, respectively, will be specified below. In Eq.~\eqref{appendix:kinetic equation}, we neglected the existence of the orbital magnetic moments of electrons in bands with nontrivial geometry. The inclusion of these orbital moments affects intravalley perturbations only, which do not change neither the total number of particles, not the total energy of a Weyl node. Therefore, they are relaxed on short timescales associated with the intravalley impurity or electron-electron scattering, and lead to small effects as compared to those associated with the chiral anomaly and the chiral magnetic effects, despite the claims of the comparable effects in Refs.~\onlinecite{Meng2020conductance,Garate2020phonon}. 
Further, we have neglected the anomalous velocity associated with the effective electric field acting on the electrons, which plays no role in sound absorption. Finally, in what follows we will also disregard the usual magnetic part of the Lorentz force, $e\bm v_w\times\bm B$, which is behind the usual magneto-absorption phenomena, not treated in this work. 

In what follows, we will neglect the density of state correction factor $D_{\bm B}$ both in Eq.~\eqref{appendix:kinetic equation} and in all observables. The reason for it is that it can only provide very small corrections to various observables. We can estimate the order of these corrections from the following argument. In the semiclassical regime (small $B$-fields), magnetic field corrections to various observables must be analytic and even functions of the magnetic field, which implies that $D_{\bm B}$-related corrections are of order of $(D_{\bm B}-1)^2\sim (\omega_c/E_F)^2$, where $\omega_c=eBv^2/E_F$ is the effective cyclotron frequency. At the same time, it is evident from Eqs.~\eqref{eq:scalarnoB} and \eqref{eq:smallomegaScalarB} that the anomaly-related contribution to magneto-absorption is governed by a much larger parameter $(\omega_c/E_F)^2\tau_v/\tau$. Hence our decision to drop $D_{\bm B}$ everywhere.

To facilitate converting the kinetic equation~\eqref{appendix:kinetic equation} into a macroscopic transport equation, we represent the distribution function near a particular Weyl point as
\begin{align}\label{appendix:dfdefinition}
    f_w=f_{eq}(E_w(\pp)-(\mu_0+\mu_w))+\delta f_w, 
\end{align}
where $\mu_0$ is the equilibrium chemical potential, $\mu_w=\lambda_wu+e\phi+n_w/\nu_w$ (see Eq.~\eqref{eq:elchempot} of the main text), and $\delta f_w$ has zero average over the Fermi surface in the $w$'th valley. Note that $\lambda_wu$ and $e\phi$ cancel out from the argument of $f_{eq}(E_w(\pp)-(\mu_0+\mu_w))=f_{eq}(\varepsilon_w(\pp)-(\mu_0+n_w/\nu_w))$. That is, $f_{eq}(E_w(\pp)-(\mu_0+\mu_w))$ describes a Weyl point ``lifted'' in energy by $\lambda_wu+e\phi$, such that the occupation numbers in the momentum space change only due to a change in the doping level of a Weyl point (counted from the Weyl point energy) by $n_w/\nu_w$. 

The two parts of the distribution function~\eqref{appendix:dfdefinition} are relaxed on very different time scales: $\delta f_w$ is relaxed by fast intravalley scattering, while $f_{eq}(E_\pp-(\mu_0+\mu_w))$, with non-equilibrium valley-dependent $\mu_w$, can only be relaxed by intervalley scattering. This hierarchy of relaxation rates allows us to neglect $\delta f_w$ in the intervalley collision integral. For our purposes, it is sufficient to write phenomenological, but physically motivated expressions for the two collision integrals in the kinetic equation~\eqref{appendix:kinetic equation}. The intravalley one we will just write in the usual constant relaxation time approximation:
\begin{align}\label{appendix:intracollision}
    &I^{\rm{intra}}_w=-\frac{\delta f_w}{\tau_w},
\end{align}
where $\tau_w$ is the intravalley transport relaxation time. It is worth looking more carefully at the intervalley collision integral, which gives rise to the right hand side of Eq.~\eqref{eq:transport} of the main text. The microscopic expression for this collision integral, which essentially follows from the Fermi golden rule, involves the scattering probability between valleys $w$ and $w'$, averaged over the Fermi surfaces in the two valleys, $\overline{W}_{ww'}(E_{w}(\pp),E_{w'}(\pp'))\delta(E_w(\pp)-E_{w'}(\pp'))$. The energies in the energy-conserving $\delta$-function are given by Eq.~\eqref{appendix:totalenergy}, which includes both the intravalley kinetic energy, and the potential energy due to the acoustic perturbation. The collision integral can be written as 
\begin{widetext}
\begin{align}\label{appendix:intercollision}
    I^{\rm{inter}}_w=-\sum_{w'}\int_{\pp'} \overline{W}_{ww'}\delta(E_w(\pp)-E_{w'}(\pp'))(f_{eq}(E_w(\pp)-(\mu_0+\mu_w))-f_{eq}(E_{w'}(\pp')-(\mu_0+\mu_{w'}))). 
\end{align}
\end{widetext}
We suppressed the energy dependence of $\overline{W}_{ww'}$, since it effectively has to be evaluated at the respective Fermi surfaces. We also assume that $\overline{W}_{ww'}=\overline{W}_{w'w}$, that is, we neglect ``valley skew scattering''. Now we expand the difference of the distribution functions in small $\mu_w-\mu_{w'}\sim O(u_{ij})$, after which all quantities pertaining to scattering can be calculated for the crystal without the acoustic perturbation. We also assume the low-temperature limit, in which $-\partial_{E_w}f_{eq}(E_w-\mu_0-\mu_w)=\delta(E_w-\mu_0-\mu_w)$. After these manipulations, the intervalley collision integral becomes 
\begin{align}\label{appendix:intercollisionfinal}
    I^{\rm{inter}}_w=-\sum_{w'}\nu_{w'} \overline{W}_{ww'}\delta(E_w(\pp)-\mu_0-\mu_{w})(\mu_w-\mu_{w'}). 
\end{align}
For the purpose of the derivation of the transport equation~\eqref{eq:transport}, we will need the momentum space integral of $I^{\rm{inter}}_w$ over momenta near the $w$'th valley:
\begin{align}\label{appendix:integralofthecollisionintegral}
    \int_\pp I^{\rm{inter}}_w=-\sum_{w'}\nu_w\nu_{w'} \overline{W}_{ww'}(\mu_w-\mu_{w'}).
\end{align}
The constant relaxation time approximation for the intervalley collision integral, used in this work, is obtained from Eq.~\eqref{appendix:integralofthecollisionintegral}, if one sets $\overline{W}_{ww'}\to \overline{W}$. In this case we obtain
\begin{align}\label{appendix:finalintegralofthecollisionintegral}
    \int_\pp I^{\rm{inter}}_w=-\frac{\nu_w}{\tau_v}(\mu_w-\ool{\mu}), \quad \frac{1}{\tau_v}\equiv\sum_w\nu_w\overline{W}.
\end{align}
The weighted valley average of the nonequilibrium part of the electrochemical potential, $\ool{\mu}$, is defined in Eq.~\eqref{eq:weightedaverage} of the main text. 

Turning to the derivation of the macroscopic transport equation, we first integrate the kinetic equation~\eqref{appendix:kinetic equation} over the momentum near the $w$'th valley, and restrict ourselves to the linear response to the acoustic wave to obtain 
\begin{widetext}
\begin{align}\label{appendix:transportzigote}
    \partial_t n_w+\bm\nabla_\rr\int_\pp\bm v_w \delta f_w +\frac{e}{4\pi^2}\eta_w\bm B\cdot\bm\nabla_\rr\frac{n_w}{\nu_w}+\frac{e}{4\pi^2}\eta_w\bm B\cdot\bm\nabla_\rr(\lambda_wu+e\phi)=
    -\frac{\nu_w}{\tau_v}\left(\mu_w-\ool \mu\right).
\end{align}
\end{widetext}
In the transport Eq.~\eqref{appendix:transportzigote}, the second term on the left hand side is the divergence of the standard intravalley Drude (number, rather than electric) current, $\bm j_w/e$, driven by the gradient of the electrochemical potential. The expression for this current is easily obtained from the odd in momentum part of the kinetic equation, combined with the expression for the intravalley collision integral, Eq.~\eqref{appendix:intracollision}. In the diffusive regime, neglecting $\omega$ and $qv_w$ as compared to $1/\tau_w$, this yields 
\begin{align}\label{appendix:deltaf}
    \delta f_w=-\tau_w \bm v_w \left(e\bm{\mathcal{E}}_w-\nabla_\rr\frac{n_w}{\nu_w}\right)\partial_\varepsilon f_{eq}(\varepsilon_w-\mu_0),
\end{align}
and leads to $\bm j_w/e=-\nu_w D_w\nabla_\rr\mu_w$, if we neglect the tensorial nature of the diffusion constant for simplicity. Then $\nabla_\rr\cdot\bm j_w/e$ corresponds to the second term on the left hand side of Eq.~\eqref{eq:transport}. The third and forth terms on the left hand side of Eq.~\eqref{appendix:transportzigote} are specific to Weyl materials, and represent the divergence of the CME current due to a chemical potential change of $n_w/\nu_w$ in valley $w$, and the rate of change of the particle density in valley $w$ due to the chiral anomaly. To obtain these terms, we performed the appropriate momentum-space integrals that involve the Berry curvature, which are standard by now. Their detailed evaluation can be found in Ref.~\onlinecite{MaPesin2015}. Taken together, the CME and chiral anomaly related terms in Eq.~\eqref{appendix:transportzigote} combine to yield the third term in Eq.~\eqref{eq:transport} of the main text. It is noteworthy that this term involves only the total electrochemical potential of the valley\cite{ParameswaranPesin2014}. Finally, on the right hand side of Eq.~\eqref{appendix:transportzigote}, we took into account that $\int_\pp I_w^{\rm{intra}}=0$, since intravalley scattering conserves the number of particles near the corresponding valley, and used Eq.~\eqref{appendix:finalintegralofthecollisionintegral} for $\int_\pp I_w^{\rm{inter}}$, which coincides with the right hand side of Eq.~\eqref{eq:transport}, and is further discussed in the main text. This concludes our presentation as far as the transport equation Eq.~\eqref{eq:transport} of the main text is concerned.

To derive the expression for the entropy production (Eq.~\eqref{eq:entropy_production} of the main text), we start with the combinatorial expression for the electronic part of the entropy of the system:
\begin{align}
    S=-\sum_{w}\int_\rr\int_\pp (f_w\ln(f_w)+(1-f_w)\ln(1-f_w)),
\end{align}
where the space, time, and momentum dependence of $f_w(\rr,\pp,t)$ have been suppressed for brevity. 
Recall that we have been systematically neglecting the density of states factor $D_{\bm B}$, so it does not appear in the expression for the entropy either. Differentiating the expression for the entropy with respect to time, we obtain
\begin{align}\label{appendix:Sdot}
    \dot{S}=-\sum_{w}\int_\rr\int_\pp \partial_t f_w\ln\frac{f_w}{1-f_w}.
\end{align}
Further on, we substitute the expression for $\partial_t f_w$ from the kinetic equation, and note that the terms from its left hand side only contribute to the entropy redistribution in real space, which is accomplished in each valley by the heat current
\begin{align}
    \bm j^{\rm{heat}}_w=\int_\pp \bm v_w(\pp)(\varepsilon_w(\pp)-\mu^c_{w}) \delta f_w +e\eta_w\frac{T^2}{12}\bm B,
\end{align}
$\mu^c_{w}$ is the chemical (rather than electrochemical) potential in valley $w$, while $\varepsilon_w(\pp)$ is defined in Eq.~\eqref{appendix:totalenergy}. The first term in the above equation has the obvious meaning of the usual heat current of quasiparticles, while the second one is a topological contribution analogous to the chiral magnetic effect for the charge current. These convective terms do not lead to net entropy production in the sample. 

The only non-zero contribution to the entropy production comes from the collision integrals, both intra- and intervalley. To obtain the expression for the entropy production due to intravalley diffusion, we expand the log in Eq.~\eqref{appendix:Sdot} to linear order in $\delta f_w$, and use the expression for $I^{\rm{intra}}_w$ in place of $\partial_t f_w$, as well as Eq.~\eqref{appendix:deltaf} for $\delta f_w$. This immediately leads to the first -- diffusive entropy production --  term in the right hand side of Eq.~\eqref{eq:entropy_production}. (Just like in the transport equation, we neglect the tensorial nature of the diffusion coefficient.) Since the result is just an expression for the intravalley Joule heating driven by the gradient of the electrochemical potential, we do not go into further details of its derivation. Instead, we focus on intervalley scattering contribution to the entropy production. To this end we use Eq.~\eqref{appendix:intercollision} for $I^{\rm{inter}}_w$ in place of $\partial_t f_w$ in Eq.~\eqref{appendix:Sdot}, neglect $\delta f_w$ under the logarithm, assume the low-temperature limit, and neglect the energy dependence of the scattering rates and the densities of states. This way we obtain 
\begin{widetext}
\begin{align}
    T\dot S^{\rm{inter}}=\int_\rr\sum_{ww'}\nu_w\nu_{w'} \overline{W}_{ww'}\int^{\mu_0+\mu_{w'}}_{\mu_0+\mu_{w}}dE_w(E_w(\pp)-\mu_0-\mu_w)
    =\frac12\int_\rr \sum_{ww'}\nu_w\nu_{w'} \overline{W}_{ww'}(\mu_w-\mu_{w'})^2. 
\end{align}
\end{widetext}
If we now make the same assumption $\overline{W}_{ww'}\to \overline{W}$ as in deriving the transport equation, and note that $\sum_w\nu_w\mu_w=\sum\nu_w\ool{\mu}$, we will arrive at the second term on the right hand side of Eq.~\eqref{eq:entropy_production}.
\bibliography{attenuation_references}

\begin{thebibliography}{74}
\bibitem [{mag()}]{magnitudes_of_deformpotentials}%
  \BibitemOpen
  \href@noop {} {}\bibinfo {note} {A rough estimate of the magnitudes of scalar
  and vectors parts of the deformation potential can be obtained as follows.
  For an extreme deformation of order of the lattice constant on the scale of
  the lattice constant, $u_{ij}\sim 1$, the scalar and vector parts of the
  deformation potential should lead to a displacement of a Weyl node by the
  full bandwidth, or the full Brillouin zone, respectively. This implies that
  $\lambda\sim W$ and $\xi\sim v/a$, where $W$ is the bandwidth of the relevant
  band, and $a$ is the lattice constant. Given that typically $W\sim v/a$, we
  arrive at the conclusion that $\lambda$ and $\xi$ are similar in
  magnitude.}\BibitemShut {Stop}%
\end{thebibliography}

\end{document}